\documentclass[aps,prl,manuscript,superscriptaddress]{revtex4-2}
\usepackage{graphicx,wrapfig}
\usepackage{fancyhdr}
\usepackage[colorlinks,citecolor = blue,linkcolor=blue]{hyperref}
\usepackage{ifthen}
\usepackage{color}
\usepackage{epsfig}
\usepackage{dcolumn}
\usepackage{float}
\usepackage{mathtools}
\usepackage{hyperref}
\usepackage{comment}
\usepackage{bbm}
\usepackage{amsmath}

\newcommand{\TNSe}{{Ta$_2$NiSe$_5$}}
\newcommand{\TNS}{{Ta$_2$NiS$_5$}}
\newcommand{\TNSS}{{Ta$_2$Ni(Se,S)$_5$}}

\begin{document}

\title{Anomalous excitonic phase diagram in band-gap-tuned Ta$_2$Ni(Se,S)$_5$}

\author{Cheng Chen}
\email{These authors contributed equally to this work}
\affiliation{Department of Physics, University of Oxford, Oxford, OX1 3PU, United Kingdom}
\affiliation{Department of Applied Physics, Yale University, New Haven, Connecticut 06511, USA}

\author{Weichen Tang} 
\email{These authors contributed equally to this work}
\affiliation{Physics Department, University of California, Berkeley, California 94720, USA}
\affiliation{Materials Science Division, Lawrence Berkeley National Lab, Berkeley, California 94720, USA}

\author{Xiang Chen} 
\affiliation{Physics Department, University of California, Berkeley, California 94720, USA}
\affiliation{Materials Science Division, Lawrence Berkeley National Lab, Berkeley, California 94720, USA}

\author{Zhibo Kang}
\affiliation{Department of Applied Physics, Yale University, New Haven, Connecticut 06511, USA}

\author{Shuhan Ding} 
\affiliation{Department of Physics and Astronomy, Clemson University, Clemson, South Carolina 29631, USA}

\author{Kirsty Scott} 
\affiliation{Department of Physics, Yale University, New Haven, Connecticut 06511, USA}

\author{Siqi Wang} 
\affiliation{Department of Applied Physics, Yale University, New Haven, Connecticut 06511, USA}

\author{Zhenglu Li} 
\affiliation{Physics Department, University of California, Berkeley, California 94720, USA}
\affiliation{Materials Science Division, Lawrence Berkeley National Lab, Berkeley, California 94720, USA}

\author{Jacob P.C. Ruff}
\affiliation{Cornell High Energy Synchrotron Source, Cornell University, Ithaca, New York 14853, USA}

\author{Makoto Hashimoto}
\affiliation{Stanford Synchrotron Radiation Lightsource, SLAC National Accelerator Laboratory, Menlo Park, California 94025, USA}

\author{Dong-Hui~Lu}
\affiliation{Stanford Synchrotron Radiation Lightsource, SLAC National Accelerator Laboratory, Menlo Park, California 94025, USA}

\author{Chris Jozwiak}
\affiliation{Advanced Light Source, Lawrence Berkeley National Laboratory, Berkeley, California 94720, USA}

\author{Aaron Bostwick}
\affiliation{Advanced Light Source, Lawrence Berkeley National Laboratory, Berkeley, California 94720, USA}

\author{Eli Rotenberg}
\affiliation{Advanced Light Source, Lawrence Berkeley National Laboratory, Berkeley, California 94720, USA}

\author{Eduardo H. da Silva Neto}
\affiliation{Department of Physics, Yale University, New Haven, Connecticut 06511, USA}

\author{Robert J. Birgeneau}
\affiliation{Physics Department, University of California, Berkeley, California 94720, USA}
\affiliation{Materials Science Division, Lawrence Berkeley National Lab, Berkeley, California 94720, USA}
\affiliation{Department of Materials Science and Engineering, University of California, Berkeley, California 94720, USA}

\author{Yulin Chen}
\affiliation{Department of Physics, University of Oxford, Oxford, OX1 3PU, United Kingdom}

\author{Steven G. Louie} 
\email{sglouie@berkeley.edu}
\affiliation{Physics Department, University of California, Berkeley, California 94720, USA}
\affiliation{Materials Science Division, Lawrence Berkeley National Lab, Berkeley, California 94720, USA}

\author{Yao Wang} 
\email{yao.wang@emory.edu}
\affiliation{Department of Physics and Astronomy, Clemson University, Clemson, South Carolina 29631, USA}
\affiliation{Department of Chemistry, Emory University, Atlanta, GA 30322, USA}

\author{Yu He}
\email{yu.he@yale.edu}
\affiliation{Department of Applied Physics, Yale University, New Haven, Connecticut 06511, USA}
\affiliation{Physics Department, University of California, Berkeley, California 94720, USA}
\affiliation{Materials Science Division, Lawrence Berkeley National Lab, Berkeley, California 94720, USA}

\date{\today}

\begin{abstract}
During a band-gap-tuned semimetal-to-semiconductor transition, Coulomb attraction between electrons and holes can cause spontaneously formed excitons near the zero-band-gap point, or the Lifshitz transition point. This has become an important route to realize bulk excitonic insulators -- an insulating ground state distinct from single-particle band insulators. How this route manifests from weak to strong coupling is not clear. In this work, using angle-resolved photoemission spectroscopy (ARPES) and high-resolution synchrotron x-ray diffraction (XRD), we investigate the broken symmetry state across the semimetal-to-semiconductor transition in a leading bulk excitonic insulator candidate system \TNSS. A broken symmetry phase is found to be continuously suppressed from the semimetal side to the semiconductor side, contradicting the anticipated maximal excitonic instability around the Lifshitz transition. Bolstered by first-principles and model calculations, we find strong interband electron-phonon coupling to play a crucial role in the enhanced symmetry breaking on the semimetal side of the phase diagram. Our results not only provide insight into the longstanding debate of the nature of intertwined orders in \TNSe, but also establish a basis for exploring band-gap-tuned structural and electronic instabilities in strongly coupled systems.
\end{abstract}

\maketitle

\begin{figure*}
\centering
\includegraphics[width= 15 cm]{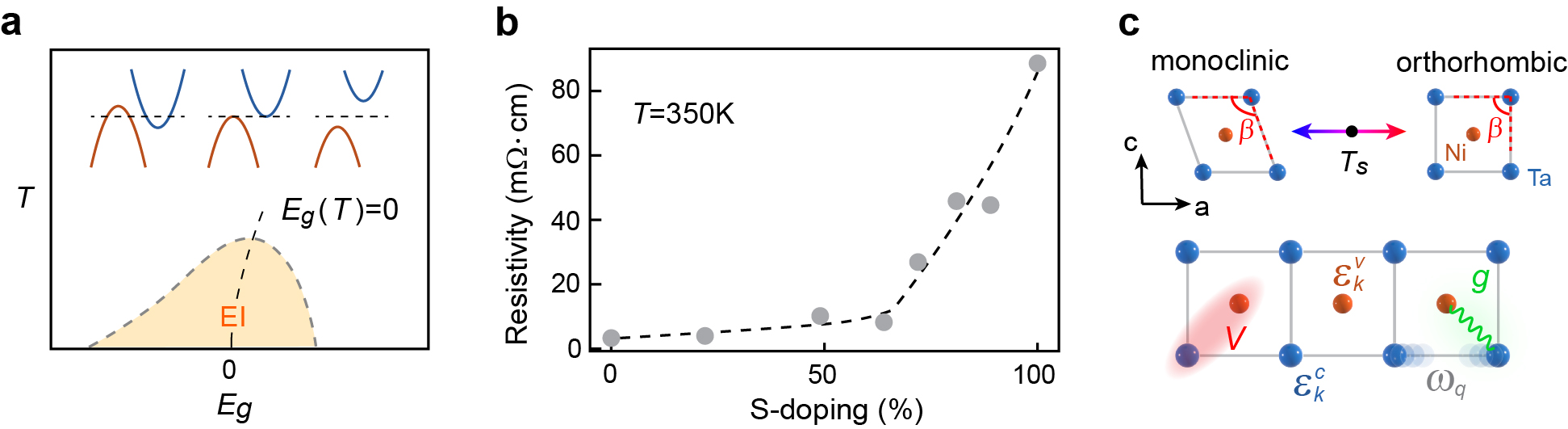}\vspace{-3mm}
\caption{\textbf{Band-gap-tuned semimetal-to-semiconductor transition in narrow band gap systems.} \textbf{a} Archetypal phase diagrams in indirect band gap controlled excitonic insulator (EI) across the semimetal-to-semiconductor transition. \textbf{b} Normal state electrical resistivity evolution with S-doping in \TNSS. \textbf{c} Structure motif of quasi-1D \TNSS~system below and above transition temperature $T_{s}$ (upper panel). The lattice distortion in the monoclinic phase is exaggerated. Symbolic sketch of a minimal lattice model of the system used to describe the low-energy interactions in the system (lower panel). $\epsilon_k^c$ and $\epsilon_k^v$ are kinetic energies of conduction and valence electrons. $\omega_q$ is the chain-shearing $B_{2g}$ phonon frequency. $g$ is the interband electron-phonon coupling vertex. $V$ is the electron-hole Coulomb interaction potential.
}
\label{fig:Fig1}
\end{figure*}

When the indirect band gap of a material is continuously reduced from positive to negative, band theory predicts a semiconductor-to-semimetal Lifshitz transition. However, such a transition may never be reached in the presence of electron-hole Coulomb attraction. It is postulated that poor screening in low carrier density semimetals and strong electron-hole binding in narrow gap semiconductors can both result in the spontaneous formation of excitons\,\cite{mott1961transition, knox1983introduction}. The condensation of excitons results in an insulating ground state, the so-called ''excitonic insulator'' (EI), which is separated from the normal state by a dome-shaped phase boundary peaked around the Lifshitz point (see Fig.~\ref{fig:Fig1}\textbf{a})\,\cite{keldysh1965possible}. This simple setup became a primary platform to search for condensed excitons and other exotic correlated phases, as well as potential solid-state BCS-BEC crossover phenomena\,\cite{halperin1968excitonic,fedders1966itinerant,arrott1965first,holger2006possibility}. For example, adding electronic anisotropy can further narrow the EI dome due to reduced Coulomb attraction\,\cite{zittartz1967anisotropy,kozlov1965metal}, and sufficiently heavy holes may cause Wigner crystallization\,\cite{halperin1968possible}. Moreover, band-gap-tuned BCS-BEC crossover has been recently suggested in 2D moir\'e systems\,\cite{kim2021spectroscopic}. However, interband electron-phonon interaction can also lead to a gapped charge-density-wave state that occurs concurrently with the exciton condensation\,\cite{halperin1968excitonic}. It remains unclear how the excitonic instability and the density-wave instability interact with each other, and whether it is possible at all to distinguish them. Since no super-transport (charge or heat) is associated with the excitonic insulator phase, and its analogy to BCS superconductivity is only formal\,\cite{halperin1968possible}, an effective experimental method to distinguish the Coulomb and lattice channels is to identify a tuning parameter that gives differentiating predictions of the phase diagram.

Band-gap-controlled materials that undergo semimetal-to-semiconductor transitions provide an ideal platform to investigate this issue. However, the experimental realization of such a transition in bulk systems has been challenging. Divalent metals under hydrostatic pressure and uniaxial strain were first proposed as promising candidates\,\cite{jerome1967excitonic}, but experimental evidence was indirect and scarce without clear ways to separate Coulomb and electron-phonon interaction effects. Recent experiments suggest a potential exciton condensation in 1\textit{T}-TiSe$_2$\,\cite{kogar2017signatures}, where the band gap tunability is limited. Quasi-one-dimensional (quasi-1D) ternary chalcogenide Ta$_2$NiSe$_5$ has recently emerged as another leading EI candidate, exhibiting an orthorhombic-to-monoclinic structural transition at $T_s=$ 329~K, concomitant with a putative exciton condensation\,\cite{di1986physical,lu2017zero}. The system consists of parallel quasi-1D Ta and Ni chains (see Fig.~\ref{fig:Fig1}\textbf{c}), where long-lived electron-hole pairs are supposed to dwell on Ta 5$d_{x^2-y^2}$ and Ni 3$d_{xy}$ orbitals. On the one hand, the sizable ground-state single-particle gap\,\cite{lee2019strong,wakisaka2009excitonic,wakisaka2012photoemission,seki2014excitonic}, minute structural distortion\,\cite{di1986physical,chen2022lattice}, and a ``dome-like'' temperature-pressure phase diagram\,\cite{lu2017zero,halperin1968possible} are all ostensibly consistent with the predictions in a band-gap-tuned EI. On the other hand, strong electron-phonon coupling effects are also found to coexist with correlation effects\,\cite{mazza2020nature,baldini2023spontaneous,chen2022lattice,watson2020band, bretscher2021ultrafast, yan2019strong, volkov2021critical,volkov2021failed,kim2021direct,larkin2017giant,larkin2018infrared,werdehausen2018coherent,bretscher2021imaging}. With the presence of both interactions, the band topology alone no longer sufficiently dictates the excitonic insulator phase region, and it is unclear how the low-temperature broken symmetry state would evolve across the Lifshitz transition point. In \TNSS, this question can be addressed since the value of the direct band gap can be effectively tuned by the isovalent sulfur substitution of selenium\,\cite{lu2017zero, Tanusree2022electronic}. However, the phase diagram of \TNSS -- especially where the broken symmetry phase ends and where the normal state crosses from a gapless to a gapped electronic structure with S-doping -- has seen vastly contradicting results~\cite{di1986physical,lu2017zero,volkov2021failed}, mainly due to the lack of direct probes of the single-particle band gap and the order parameter of the broken symmetry.  A general impact of S-substitution on the low-energy electronic structure can be revealed through the normal-state (defined here as the high-temperature symmetric structure phase) resistivity measured along the chain direction (see Fig.~\ref{fig:Fig1}\textbf{b}), which slowly increases until an abrupt upturn around $70\%$ doping. Such an upturn indicates a potential Lifshitz transition on the normal-state band structures. Hence, S-doped \TNSe\ constitutes an ideal platform to study band-gap-tuned semimetal-to-semiconductor transition in an electron-phonon coupled correlated system.

\begin{figure*}
\centering
\includegraphics[width= 12 cm]{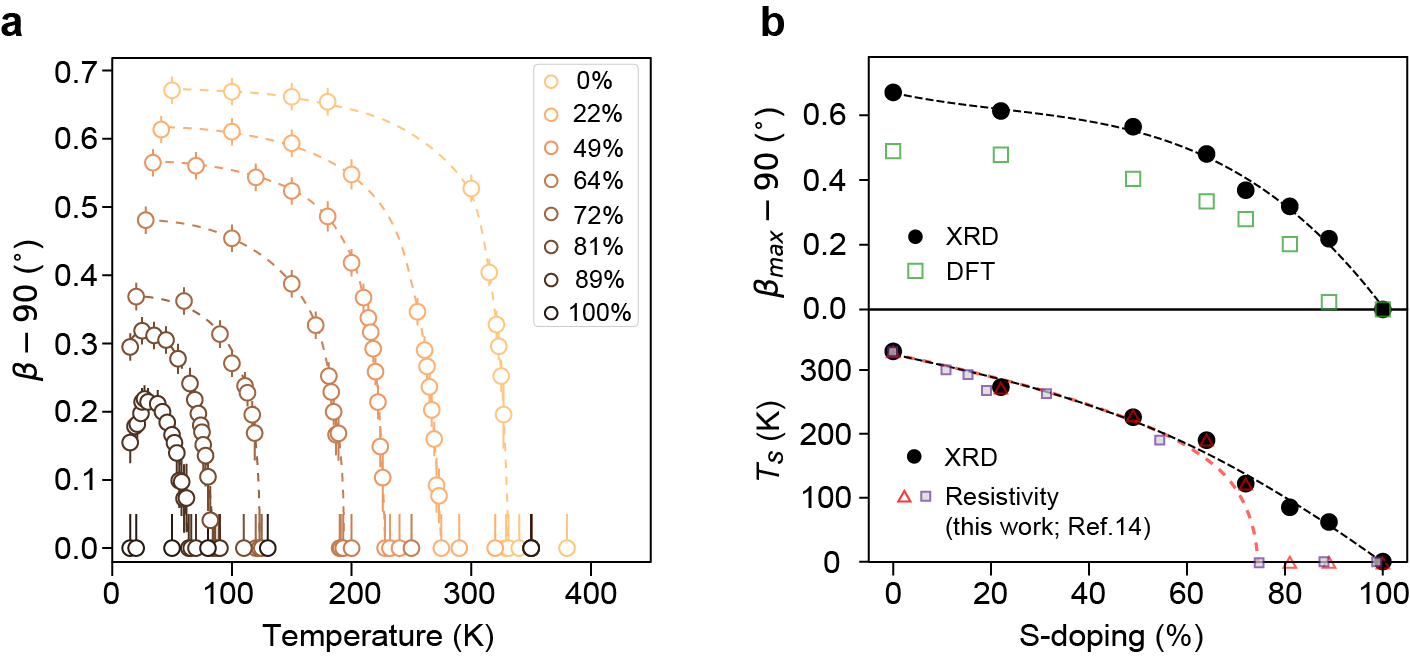}\vspace{-3mm}
\caption{\textbf{Structural order parameter of \TNSS.} \textbf{a} Change of the inter- and intra-chain Ta-Ta bond angle $\beta - 90^\circ$ as a function of temperature and S-doping. \textbf{b} Top: experimental and DFT calculated $\beta - 90^\circ$ at zero temperature. Bottom: structural transition temperature $T_s$ as a function of S-doping level, determined from XRD and resistivity measurement.}
\label{fig:Fig2}
\end{figure*}

\begin{figure*}
\centering
\includegraphics[width= 16 cm]{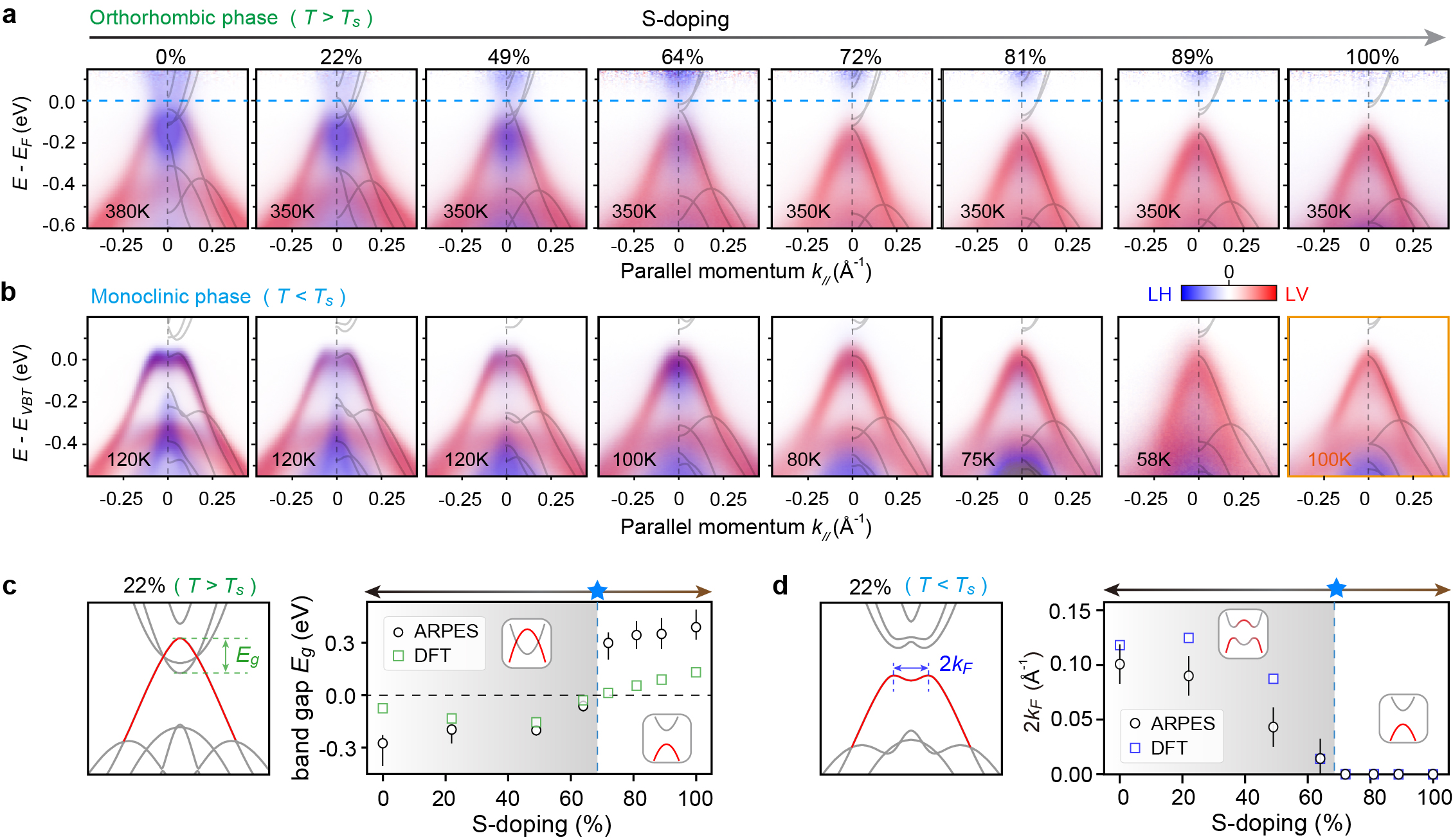}\vspace{-3mm}
\caption{\textbf{Electronic Lifshitz transition in \TNSS.} ARPES spectra along $X-\Gamma-X$ direction for samples with different S-doping levels. \textbf{a} High-temperature orthorhombic phase. \textbf{b} Low-temperature monoclinic phase (except for 100\% S-doping, where no structural transition was evidenced down to 10 K). Spectra from linear horizontal (LH, blue) and linear vertical (LV, red) polarization of incident light are overlaid, highlighting Ta 5$d_{x^2-y^2}$ conduction band and Ni 3$d_{xy}$ valence bands respectively. VBT - valence band top. \textbf{c} DFT calculation of 22\% S-doped compound in high-temperature orthorhombic phase, and the evolution of $E_g$ extracted from DFT calculation and photoemission spectra as a function of S-doping. \textbf{d} DFT calculation of 22\% S-doped compound in low-temperature monoclinic phase, and the evolution of $2k_F$ extracted from DFT calculation and photoemission spectra as a function of S-doping.}
\label{fig:Fig3}
\end{figure*}

We first track the evolution of the broken symmetry phase boundary in \TNSS with direct measurements of the structural order parameter. In bulk excitonic insulators, exciton condensation is always tied to a simultaneous lattice distortion~\cite{halperin1968possible}. Employing high-resolution single-crystal synchrotron XRD, Fig.~\ref{fig:Fig2}\textbf{a} characterizes the temperature- and S-substitution-dependent evolution of the lattice order parameter $\beta$ (also see Supplementary Note 1), which is defined as the angle between the lattice $a$ and $c$ axes shown in Fig.~\ref{fig:Fig1}\textbf{c}. We find the structural phase transition remains second-order throughout the entire S-doping range. The ground-state order parameter $\beta(T=0)$ monotonically decreases with increasing S-doping (Fig.~\ref{fig:Fig2}\textbf{b}), and the structural transition temperature only approaches zero near full S-substitution (\TNS) (bottom panel in Fig.~\ref{fig:Fig2}\textbf{b}). This is notably different from the electronic phase diagram inferred from resistivity (see Fig.~\ref{fig:Fig2}\textbf{b} and Supplementary Note 1), which instead suggests an abrupt collapse of the broken symmetry phase at an intermediate S-doping level\,\cite{lu2017zero}. Therefore, it is imperative to obtain a direct electronic structure view to determine the normal state semimetal-to-semiconductor transition point $p_L$ along the S-doping axis.

\begin{figure*}
\centering
\includegraphics[width= 17cm]{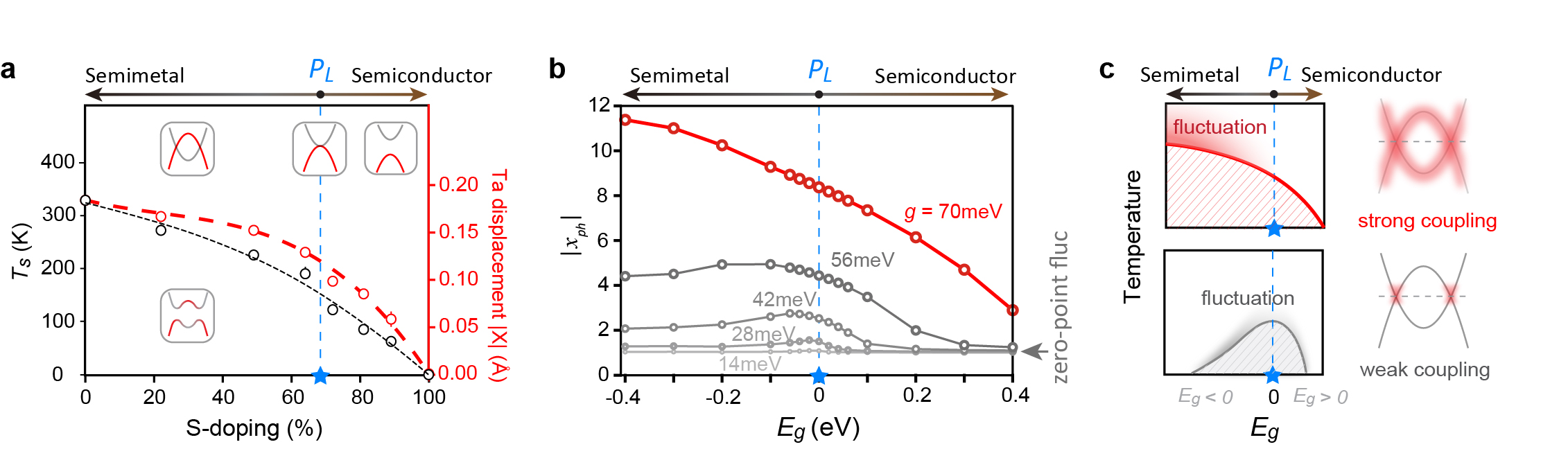}\vspace{-7mm}
\caption{\textbf{Band gap tuned phase diagram with electron-phonon coupling.} \textbf{a} Evolution of structural transition temperature $T_s$ and the maximum Ta atom displacement $|X|$ as a function of S-doping in \TNSS. The electronic band structures within different regions are illustrated in the insets. \textbf{b} The calculated average lattice displacement $|x_{\rm ph}|=\sqrt{\langle \hat{x}_{\rm ph}^2\rangle}$ as a function of the band gap $E_g$ at various electron-phonon coupling strengths. In the quantized phonon model, the zero-point fluctuation gives a finite $|x_{\rm ph}|=1$ in the decoupled limit. \textbf{c} Sketched phase diagrams of band-gap-tuned semimetal-to-semiconductor transition in the strong (top) and weak (bottom) coupling limits. Solid lines mark the phase boundaries, above which the shades show the order fluctuations\,\cite{chen2022lattice}. Red shade on the bands illustrates the electronic states that participate in the phase transition. The horizontal dashed lines denote $E_F$.}
\label{fig:Fig4}
\end{figure*}

Previous studies have reported conflicting values of $p_L$ from resistivity, Raman measurements\,\cite{lu2017zero,volkov2021failed}.  Here, we conduct high-resolution ARPES measurement, which is a direct probe of electronic structure, to reveal the evolution of band overlap/gap of \TNSS~ across the semimetal-semiconductor phase transition. The high statistics of the measurements enable electronic state restoration up to 4.5\,k$_B$T above E$_F$ (see \textbf{Methods}), and the orbital selectivity through polarized X-ray beam, allows the direct comparison with density functional theory (DFT) calculations, both of which help us precisely locate $p_L$. Figure~\ref{fig:Fig3} shows the photoemission spectra in both the normal and the broken symmetry state (also see  Supplementary Note 1). As the S-doping level increases, the Ta 5$d_{x^2-y^2}$ orbital component (blue) gradually withdraws from the valence band top, which becomes dominated by the Ni 3$d_{xy}$ orbital (red). Pseudogap states, the strong depletion of spectra intensity around Fermi level \cite{seki2014excitonic,chen2022lattice}, are also observed, which account for the high-temperature insulating behavior in these semimetallic samples (see Ref.\onlinecite{chen2022lattice} for detailed description). Beyond $p_L$ ($\sim70\%$ S-doping), the normal-state conduction and valence bands are completely separated in energy (see Fig.~\ref{fig:Fig3}\textbf{a}) and, simultaneously in the broken-symmetry-state, the Ta orbital component largely disappears from the valence band top (see Fig.~\ref{fig:Fig3}\textbf{b}).  Fitting the normal-state band dispersions (see  Supplementary Note 2 for details) quantifies the band gap (or band overlap in the case of negative values), which changes sign across $p_L$ (Fig.~\ref{fig:Fig3}\textbf{c}). In addition, revealed through the ``M-shape'' dispersion of the valence band top in the broken symmetry state, the band back-bending momentum $k_F$ also approaches $\Gamma$ near $p_L$ (Fig.~\ref{fig:Fig3}\textbf{d}). Both results are consistent with a concurrent normal state semimetal-to-semiconductor transition. Our DFT results further confirm such a Lifshitz transition near $p_L$ via the band-gap-sign flip and the $k_F$ reduction at the single-particle level, without involving \textit{ad hoc} electron-hole Coulomb attraction (see Fig.~\ref{fig:Fig3}\textbf{c-d} and  Supplementary Note 3).


The ARPES measurement of the semimetal-to-semiconductor transition doping is consistent with that deduced from resistivity measurement (Fig.~\ref{fig:Fig1}\textbf{b} and \ref{fig:Fig2}\textbf{b}). However, the structural phase transition temperature $T_s$ and the structural order parameter (approximated as $\beta_{\text{max}}$ - 90$^{\circ}$ ) are found to decrease monotonically with increasing S-doping through our XRD measurements (Fig. \ref{fig:Fig2}\textbf{b}). This poses a marked departure from the archetypal band-gap-tuned excitonic insulator phase diagram, where the excitonic instability maximizes around $p_L$ (Fig.~\ref{fig:Fig1}\textbf{a}). To highlight this contrast, Figure~\ref{fig:Fig4}\textbf{a} shows the evolution of both the measured structural transition temperature $T_s$ and the experimentally determined Ta atom displacement $|X|$ in \TNSS. Both $|X|$ and $T_s$ appear insensitive to the system crossing $p_L$, which leads to considerations beyond weak electron-hole Coulomb interaction. One possibility is a heavy-hole Wigner crystal phase\,\cite{halperin1968possible}; but the electron and hole effective masses are both near the free electron mass in \TNSS\,\cite{Tanusree2022electronic}. Alternatively, strong Coulomb interaction may push the phase boundary towards the semiconductor side~\cite{sugimoto2018strong}, but the Hartree component will undo this effect\cite{chen2022lattice} (see also Supplementary Note 4).
Considering the fact that strong electron-phonon coupling effect is revealed in both equilibrium and ultrafast pump-probe experiments in \TNSS\,\cite{mazza2020nature,baldini2023spontaneous,chen2022lattice,watson2020band, bretscher2021ultrafast, yan2019strong, volkov2021critical,volkov2021failed,kim2021direct}, we hereby investigate an experimentally informed minimal model with strong electron-phonon coupling, shown in Fig.~\ref{fig:Fig1}\textbf{c} and Eq.~\eqref{eq:eqn1} of \textbf{Methods}. This model contains both electron-hole Coulomb interactions and an interband electron-phonon coupling, which can well describe both the spectral properties of \TNSe~ below and above $T_s$\,\cite{chen2022lattice}. To examine the impact of strong electron-phonon coupling on the band-gap-tuned transition as a matter-of-principle question, we first set the inter-band Coulomb interaction to zero. The role of this Coulomb interaction primarily lies in a Hartree shift between two bands and can be mapped to a correction of $E_g$ (see detailed discussions in Supplementary Note 4). The average phonon displacement $|x_{\rm ph}|$ at $T=0$ is employed to approximate the order parameters $|X|$ measured in the experiment (Fig.~\ref{fig:Fig4}\textbf{a}). With the quantized phonon modes, a zero-point fluctuation leads to the zero-coupling baseline at $|x_{\rm ph}|=1$. 

Solved with exact diagonalization, $|x_{\rm ph}|$ for various $E_g$ and coupling strengths are shown in Fig.~\ref{fig:Fig4}\textbf{b}. In the weak-coupling case (e.g.~$g=14$~meV), the lattice order parameter displays a dome-like structure similar to the EI phase diagram derived based on Fermi-surface instability in Fig.~\ref{fig:Fig1}\textbf{a}. With increasing coupling strengths, the dome gradually evolves into a monotonic crossover. This evolution exists not only for the many-body measure of fluctuations, but also for order parameters when symmetry breaking is allowed (see Supplementary Note 4). The monotonic crossover driven by strong, non-perturbative interactions reflects the irrelevance of Fermi-surface instability (top panel in Fig.~\ref{fig:Fig4}\textbf{c}). The coupling strength in \TNSe, estimated from both ARPES spectral fitting and first-principles deformation potentials\,\cite{chen2022lattice}, is 50$\sim$60\,meV, placing it in the strong-coupling regime. While more complicated models with multiple bands may be necessary to quantitatively reproduce the order parameter evolution, strong electron-phonon coupling qualitatively describes the evolution of the broken-symmetry phase across $p_L$ in experiments. Moreover, strong coupling also expands the momentum range over which the lattice becomes unstable\,\cite{zhu2015classification}, opening up the opportunity for a $\mathbf{q}=0$ transition when 2$k_F$ is small. 

We demonstrate that strong electron-phonon interaction can greatly enhance the structural symmetry breaking in a semimetal, and the evolution of the broken-symmetry phase boundary can be indifferent to the semimetal-to-semiconductor transition. On the specific debate over the nature of the symmetry breaking in \TNSS~systems, our results suggest a substantial deviation from the electron-hole Coulomb attraction-based EI phase (bottom panel in Fig.~\ref{fig:Fig4}\textbf{c}). Instead, a strong electron-phonon-coupling induced structural transition can explain the observed symmetry breaking. This conclusion is consistent with recent experimental evidence of the crucial role of phonons in the equilibrium and nonequilibrium properties of  \TNSe\,\cite{baldini2023spontaneous, watson2020band}. Meanwhile, it has been suggested recently that additional Coulomb and acoustic-phonon effects can also cooperatively contribute to the transition\,\cite{volkov2021failed}. On the high-temperature fluctuations, previous investigations have identified a normal-state spectral ``pseudogap'', characterized by an anomalous spectral weight accumulation near the conduction band bottom. This phenomenon is interpreted as the signature of preformed excitons due to strong electronic correlation or electron coupling to lattice fluctuations\,\cite{fukutani2021detecting,chen2022lattice}. We find that such a spectral anomaly ubiquitously exists in the normal state on the semimetal side of \TNSS, but rapidly disappears on the semiconductor side (Fig.~\ref{fig:Fig3}\textbf{a}). Intriguingly, the boundary of the pseudogap, or preformed excitons, is almost temperature-independent, similar to the abruptly terminated pseudogap in hole-doped superconducting cuprates\,\cite{he2018rapid,chen2019incoherent}. Last but not least, the heavily S-substituted \TNSS\ system provides a promising platform to study both solid state BCS-BEC crossover phenomenon~\cite{hogler2010spectral} and critical fluctuations of low-energy phonons. Specifically, a potential competing order is observed in Fig.~\ref{fig:Fig2}\textbf{a} for S-dopings above $p_L$, similar to the situation in iron-based high $T_c$ superconductors, where multiple orders are found competing with each other at low temperatures, including nematicity, spin-density wave, and superconductivity\cite{yi2014dynamic, fernandes2014drives}.  


\section*{Methods}
\subsection*{Sample synthesis and resistivity measurements}
Single crystals of \TNSS~ were grown by the chemical vapor transfer method with iodine (I$_2$) as the transport agent. Starting materials composed of Ta powder (99.99\%), Ni powder (99.99\%), Se powder (99.99\%), and S powder (99.99\%) with a nominal molar ratio 2:1:5(1-$x$):5$x$ ($x$ is the nominal doping level of S) were fully ground and mixed together inside the glovebox. About 50 mg of iodine (I$_2$) was added to the mixture of the starting powder. The mixture was then vacuumed, back-filled with 1/3 Argon, and sealed inside a quartz tube with an inner diameter of 8 $mm$, an outer diameter of 12 $mm$, and a length of about 120 $mm$. The sealed quartz tube was placed horizontally inside a muffle furnace during the growth. The hot end reaction temperature was set to 950 $^\circ$C, and the cold end was left in the air with the temperature stabilized at 850 $^\circ$C. Long and thin single crystals were harvested by quenching the furnace in the air after one week of reaction. Excess iodine was removed from the surfaces of the crystals with ethanol. Electrical resistivity measurements were carried out on a commercial PPMS (Quantum Design) by the four-probe method with the current applied along the \textit{a}-axis of the \TNSS~ single crystals. 

\subsection*{Angle-resolved photoemission spectroscopy (ARPES)}
Synchrotron-based ARPES measurements were performed at the beamline BL5-2 of Stanford Synchrotron Radiation Laboratory (SSRL), SLAC, USA, and the BL 7.0.2 of the Advanced Light Source (ALS), USA. The samples were cleaved in situ and measured under the ultra-high vacuum below 3×10$^{-11}$ Torr. Data was collected by the R4000 and DA30L analyzers. The total energy and angle resolutions were 10 meV and 0.2$^\circ$, respectively. High statistics measurements were done with a signal-to-noise ratio of 100$\sim$150 at the Fermi level on the energy distribution curve taken at the Fermi momentum. This enables electronic structure restoration up to 4.5$\sim$5 $k_B T$ after Fermi function division\,\cite{he2021superconducting,chen2019incoherent}. Orbital content selection is based on the dipole transition matrix element effect, following the experimental geometry described in Ref.~\cite{chen2022lattice}.

\subsection*{Single crystal x-ray scattering}
Hard X-ray single-crystal diffraction is carried out at the energy of 44 keV at the beamline QM2 of the Cornell High Energy Synchrotron Source (CHESS). The needle-like sample is chosen with a typical lateral dimension of 100 microns, then mounted with GE Varnish on a rotating pin before being placed in the beam. A Pilatus 6M 2D area detector is used to collect the diffraction pattern with the sample rotated 365$^\circ$ around three different axes at a 0.1$^\circ$ step and 0.1s/frame data rate at each temperature. The full 3D intensity cube is stacked and indexed with the beamline software.

\subsection*{First-principles DFT calculation}
\textit{Ab initio} calculations are performed using the Quantum ESPRESSO package\,\cite{giannozzi2009quantum,giannozzi2017advanced}. The crystal structure relaxations are performed using the r2SCAN functional\,\cite{furness2020accurate} with a semiempirical Grimme's DFT-D2 van der Waals correction\,\cite{grimme2006semiempirical}. S-doped compounds are simulated by the virtual crystal approximation (VCA), where the pseudopotentials of Se and S atoms are linearly interpolated (i.e. mixed together) according to the chemical composition. A 30$\times$30$\times$15 \textbf{k}-mesh and a 100 Ry wavefunction energy cut-off were used. The electronic structure calculations are performed using the same r2SCAN functional with the relaxed structures.

\subsection*{Two-band model for many-body simulations}
We consider the two-band model with inter-band e-ph couplings to describe the physics in the \TNSS\ systems, whose Hamiltonian reads as\,\cite{kaneko2013orthorhombic}:
\begin{eqnarray}
\mathcal{H}(E_g) &=& \sum_{k\sigma}\varepsilon_k^c(E_g) c_{k\sigma}^\dagger c_{k\sigma} + \sum_{k\sigma}\varepsilon_k^v(E_g) f_{k\sigma}^\dagger f_{k\sigma} + V\sum_{i,\sigma,\sigma^\prime} (n_{i\sigma}^c + n_{i+1\sigma}^c) n_{i\sigma'}^f\nonumber\\
&&+ \sum_{kq\sigma} \frac{g_q}{\sqrt{N}}\left[(a_q+a_{-q}^\dagger) c_{k+q\sigma}^\dagger f_{k\sigma} + H.c.  \right]+ \sum_q \omega_q a_q^\dagger a_q 
\label{eq:eqn1}
\end{eqnarray}
where $c_{k\sigma}^\dagger$ ($c_{k\sigma}$) creates (annihilates) an electron at the conduction band (primarily Ta 5$d$) for momentum k and spin $\sigma$, with dispersion given by $\varepsilon_k^c$; and the $f_{k\sigma}^\dagger$ ($f_{k\sigma}$) creates (annihilates) an electron at the valence band (primarily Ni 3$d$), with dispersion given by $\varepsilon_k^v$ (Fig.~\ref{fig:Fig1}\textbf{c}). The $n_{i\sigma}^c$ and $n_{i\sigma}^f$ are the density operators for the conduction and valence band, respectively. We employ the band structures fitted from the \TNSe\ experiments, as reported in Ref.~\onlinecite{chen2022lattice}, which determine the valence and conduction band dispersion for a negative band gap $E_g=-0.3$\,eV. Due to the difficulty of fitting the full dispersions in the gapped \TNSS\ materials, we simplify the model using a rigid band separation controlled by $E_g$. That being said, the conduction and valence band structures for \TNSS\ read as
\begin{eqnarray}
&&\varepsilon^c_k(E_g)=3.25 - 1.8\,{{\mathrm{cos}}}(k) - 0.9\,{{\mathrm{cos}}}(2k) - 0.6\,{{\mathrm{cos}}}(3k) + E_g/2\\
&&\varepsilon^v_k(E_g)=-1.95 +1.5\,{{\mathrm{cos}}}(k) +0.3\,{{\mathrm{cos}}}(2k) +0.1\,{{\mathrm{cos}}}(3k) - E_g/2\,.
\end{eqnarray}
Here, $E_g=0$ indicates the Lifshitz transition where $\varepsilon^c_{k=0} =\varepsilon^v_{k=0}$, while $E_g=0.3$\,eV reflects the situation of \TNS.

To approximate the lattice distortion in a finite-size simulation, we employ the average displacement of phonons
\begin{eqnarray}
|x_{\rm ph}|=\sqrt{\mathrm{Tr}\left[\rho\hat{x}_{\rm ph}^2\right]}\,,
\label{eq:phononDisp}
\end{eqnarray}
where $\rho$ is the density matrix of the e-ph system and $\hat{x}_{\rm ph}=(a_0^\dagger + a_0)/N$ is the uniform phonon displacement operator\,\cite{finiteSize}. 

\bigbreak

\section*{Acknowledgements}
The authors wish to thank L. Kemper, E. Demler, D. Y. Qiu, P. J. Guo, P. A. Volkov for helpful discussions. Use of the Stanford Synchrotron Radiation Light Source, SLAC National Accelerator Laboratory, is supported by the US Department of Energy, Office of Science, Office of Basic Energy Sciences under Contract No. DE-AC02-76SF00515. This research used resources of the Advanced Light Source, a US DOE Office of Science User Facility under Contract No. DE-AC02-05CH11231. This work is based upon research conducted at the Center for High Energy X-ray Sciences (CHEXS) which is supported by the National Science Foundation under award DMR-1829070. Work at Lawrence Berkeley National Laboratory was funded by the U.S. Department of Energy, Office of Science, Office of Basic Energy Sciences, Materials Sciences and Engineering Division under Contract No. DE-AC02-05-CH11231 within the Quantum Materials Program (KC2202) which provided the numerical simulations and within the Theory of Materials Program (KC2301) which provided the DFT calculations. S.D. and Y.W. acknowledge support from U.S. Department of Energy, Office of Science, Basic Energy Sciences, under Early Career Award No.~DE-SC0022874. The electron-phonon model and DFT simulations were performed on the Frontera and Stampede2 computing system, respectively, at the Texas Advanced Computing Center. The work at Yale University is partially supported by the National Science Foundation (NSF) under DMR-2239171.

\bibliography{TNSS_main.bib}

\begin{thebibliography}{48}%
\makeatletter
\providecommand \@ifxundefined [1]{%
 \@ifx{#1\undefined}
}%
\providecommand \@ifnum [1]{%
 \ifnum #1\expandafter \@firstoftwo
 \else \expandafter \@secondoftwo
 \fi
}%
\providecommand \@ifx [1]{%
 \ifx #1\expandafter \@firstoftwo
 \else \expandafter \@secondoftwo
 \fi
}%
\providecommand \natexlab [1]{#1}%
\providecommand \enquote  [1]{``#1''}%
\providecommand \bibnamefont  [1]{#1}%
\providecommand \bibfnamefont [1]{#1}%
\providecommand \citenamefont [1]{#1}%
\providecommand \href@noop [0]{\@secondoftwo}%
\providecommand \href [0]{\begingroup \@sanitize@url \@href}%
\providecommand \@href[1]{\@@startlink{#1}\@@href}%
\providecommand \@@href[1]{\endgroup#1\@@endlink}%
\providecommand \@sanitize@url [0]{\catcode `\\12\catcode `\$12\catcode `\&12\catcode `\#12\catcode `\^12\catcode `\_12\catcode `\%12\relax}%
\providecommand \@@startlink[1]{}%
\providecommand \@@endlink[0]{}%
\providecommand \url  [0]{\begingroup\@sanitize@url \@url }%
\providecommand \@url [1]{\endgroup\@href {#1}{\urlprefix }}%
\providecommand \urlprefix  [0]{URL }%
\providecommand \Eprint [0]{\href }%
\providecommand \doibase [0]{https://doi.org/}%
\providecommand \selectlanguage [0]{\@gobble}%
\providecommand \bibinfo  [0]{\@secondoftwo}%
\providecommand \bibfield  [0]{\@secondoftwo}%
\providecommand \translation [1]{[#1]}%
\providecommand \BibitemOpen [0]{}%
\providecommand \bibitemStop [0]{}%
\providecommand \bibitemNoStop [0]{.\EOS\space}%
\providecommand \EOS [0]{\spacefactor3000\relax}%
\providecommand \BibitemShut  [1]{\csname bibitem#1\endcsname}%
\let\auto@bib@innerbib\@empty
\bibitem [{\citenamefont {Mott}(1961)}]{mott1961transition}%
  \BibitemOpen
  \bibfield  {author} {\bibinfo {author} {\bibfnamefont {N.~F.}\ \bibnamefont {Mott}},\ }\bibfield  {title} {\bibinfo {title} {\textit{The transition to the metallic state}},\ }\href@noop {} {\bibfield  {journal} {\bibinfo  {journal} {Philos. Mag.}\ }\textbf {\bibinfo {volume} {6}},\ \bibinfo {pages} {287} (\bibinfo {year} {1961})}\BibitemShut {NoStop}%
\bibitem [{\citenamefont {Knox}(1983)}]{knox1983introduction}%
  \BibitemOpen
  \bibfield  {author} {\bibinfo {author} {\bibfnamefont {R.}~\bibnamefont {Knox}},\ }\bibfield  {title} {\bibinfo {title} {\textit{Introduction to exciton physics}},\ }\href@noop {} {\bibfield  {journal} {\bibinfo  {journal} {Collective Excitations in Solids}\ ,\ \bibinfo {pages} {183}} (\bibinfo {year} {1983})}\BibitemShut {NoStop}%
\bibitem [{\citenamefont {Keldysh}\ and\ \citenamefont {Kopaev}(1965)}]{keldysh1965possible}%
  \BibitemOpen
  \bibfield  {author} {\bibinfo {author} {\bibfnamefont {L.}~\bibnamefont {Keldysh}}\ and\ \bibinfo {author} {\bibfnamefont {Y.~V.}\ \bibnamefont {Kopaev}},\ }\bibfield  {title} {\bibinfo {title} {\textit{Possible instability of semimetallic state toward Coulomb interaction}},\ }\href@noop {} {\bibfield  {journal} {\bibinfo  {journal} {Sov. Phys. Solid State}\ }\textbf {\bibinfo {volume} {6}},\ \bibinfo {pages} {2219} (\bibinfo {year} {1965})}\BibitemShut {NoStop}%
\bibitem [{\citenamefont {Halperin}\ and\ \citenamefont {Rice}(1968{\natexlab{a}})}]{halperin1968excitonic}%
  \BibitemOpen
  \bibfield  {author} {\bibinfo {author} {\bibfnamefont {B.}~\bibnamefont {Halperin}}\ and\ \bibinfo {author} {\bibfnamefont {T.}~\bibnamefont {Rice}},\ }\bibfield  {title} {\bibinfo {title} {\textit{The excitonic state at the semiconductor-semimetal transition}},\ }\href@noop {} {\bibfield  {journal} {\bibinfo  {journal} {Solid State Phys.}\ }\textbf {\bibinfo {volume} {21}},\ \bibinfo {pages} {115} (\bibinfo {year} {1968}{\natexlab{a}})}\BibitemShut {NoStop}%
\bibitem [{\citenamefont {Fedders}\ and\ \citenamefont {Martin}(1966)}]{fedders1966itinerant}%
  \BibitemOpen
  \bibfield  {author} {\bibinfo {author} {\bibfnamefont {P.~A.}\ \bibnamefont {Fedders}}\ and\ \bibinfo {author} {\bibfnamefont {P.~C.}\ \bibnamefont {Martin}},\ }\bibfield  {title} {\bibinfo {title} {\textit{Itinerant antiferromagnetism}},\ }\href@noop {} {\bibfield  {journal} {\bibinfo  {journal} {Phys. Rev.}\ }\textbf {\bibinfo {volume} {143}},\ \bibinfo {pages} {245} (\bibinfo {year} {1966})}\BibitemShut {NoStop}%
\bibitem [{\citenamefont {Arrott}\ \emph {et~al.}(1965)\citenamefont {Arrott}, \citenamefont {Werner},\ and\ \citenamefont {Kendrick}}]{arrott1965first}%
  \BibitemOpen
  \bibfield  {author} {\bibinfo {author} {\bibfnamefont {A.}~\bibnamefont {Arrott}}, \bibinfo {author} {\bibfnamefont {S.}~\bibnamefont {Werner}},\ and\ \bibinfo {author} {\bibfnamefont {H.}~\bibnamefont {Kendrick}},\ }\bibfield  {title} {\bibinfo {title} {\textit{First-order magnetic phase change in chromium at 38.5 $^\circ${C}}},\ }\href@noop {} {\bibfield  {journal} {\bibinfo  {journal} {Phys. Rev. Lett.}\ }\textbf {\bibinfo {volume} {14}},\ \bibinfo {pages} {1022} (\bibinfo {year} {1965})}\BibitemShut {NoStop}%
\bibitem [{\citenamefont {Bronold}\ and\ \citenamefont {Fehske}(2006)}]{holger2006possibility}%
  \BibitemOpen
  \bibfield  {author} {\bibinfo {author} {\bibfnamefont {F.~X.}\ \bibnamefont {Bronold}}\ and\ \bibinfo {author} {\bibfnamefont {H.}~\bibnamefont {Fehske}},\ }\bibfield  {title} {\bibinfo {title} {\textit{Possibility of an excitonic insulator at the semiconductor-semimetal transition}},\ }\href@noop {} {\bibfield  {journal} {\bibinfo  {journal} {Phys. Rev. B}\ }\textbf {\bibinfo {volume} {74}},\ \bibinfo {pages} {165107} (\bibinfo {year} {2006})}\BibitemShut {NoStop}%
\bibitem [{\citenamefont {Zittartz}(1967)}]{zittartz1967anisotropy}%
  \BibitemOpen
  \bibfield  {author} {\bibinfo {author} {\bibfnamefont {J.}~\bibnamefont {Zittartz}},\ }\bibfield  {title} {\bibinfo {title} {\textit{Anisotropy effects in the excitonic insulator}},\ }\href@noop {} {\bibfield  {journal} {\bibinfo  {journal} {Phys. Rev.}\ }\textbf {\bibinfo {volume} {162}},\ \bibinfo {pages} {752} (\bibinfo {year} {1967})}\BibitemShut {NoStop}%
\bibitem [{\citenamefont {Kozlov}\ and\ \citenamefont {Maksimov}(1965)}]{kozlov1965metal}%
  \BibitemOpen
  \bibfield  {author} {\bibinfo {author} {\bibfnamefont {A.}~\bibnamefont {Kozlov}}\ and\ \bibinfo {author} {\bibfnamefont {L.}~\bibnamefont {Maksimov}},\ }\bibfield  {title} {\bibinfo {title} {\textit{The metal-dielectric divalent crystal phase transition}},\ }\href@noop {} {\bibfield  {journal} {\bibinfo  {journal} {Sov. Phys. JETP}\ }\textbf {\bibinfo {volume} {21}},\ \bibinfo {pages} {790} (\bibinfo {year} {1965})}\BibitemShut {NoStop}%
\bibitem [{\citenamefont {Halperin}\ and\ \citenamefont {Rice}(1968{\natexlab{b}})}]{halperin1968possible}%
  \BibitemOpen
  \bibfield  {author} {\bibinfo {author} {\bibfnamefont {B.}~\bibnamefont {Halperin}}\ and\ \bibinfo {author} {\bibfnamefont {T.}~\bibnamefont {Rice}},\ }\bibfield  {title} {\bibinfo {title} {\textit{Possible anomalies at a semimetal-semiconductor transition}},\ }\href@noop {} {\bibfield  {journal} {\bibinfo  {journal} {Rev. Mod. Phys.}\ }\textbf {\bibinfo {volume} {40}},\ \bibinfo {pages} {755} (\bibinfo {year} {1968}{\natexlab{b}})}\BibitemShut {NoStop}%
\bibitem [{\citenamefont {Kim}\ \emph {et~al.}(2021{\natexlab{a}})\citenamefont {Kim}, \citenamefont {Choi}, \citenamefont {Lewandowski}, \citenamefont {Thomson}, \citenamefont {Zhang}, \citenamefont {Polski}, \citenamefont {Watanabe}, \citenamefont {Taniguchi}, \citenamefont {Alicea},\ and\ \citenamefont {Nadj-Perge}}]{kim2021spectroscopic}%
  \BibitemOpen
  \bibfield  {author} {\bibinfo {author} {\bibfnamefont {H.}~\bibnamefont {Kim}}, \bibinfo {author} {\bibfnamefont {Y.}~\bibnamefont {Choi}}, \bibinfo {author} {\bibfnamefont {C.}~\bibnamefont {Lewandowski}}, \bibinfo {author} {\bibfnamefont {A.}~\bibnamefont {Thomson}}, \bibinfo {author} {\bibfnamefont {Y.}~\bibnamefont {Zhang}}, \bibinfo {author} {\bibfnamefont {R.}~\bibnamefont {Polski}}, \bibinfo {author} {\bibfnamefont {K.}~\bibnamefont {Watanabe}}, \bibinfo {author} {\bibfnamefont {T.}~\bibnamefont {Taniguchi}}, \bibinfo {author} {\bibfnamefont {J.}~\bibnamefont {Alicea}},\ and\ \bibinfo {author} {\bibfnamefont {S.}~\bibnamefont {Nadj-Perge}},\ }\bibfield  {title} {\bibinfo {title} {\textit{Spectroscopic signatures of strong correlations and unconventional superconductivity in twisted trilayer graphene}},\ }\href@noop {} {\bibfield  {journal} {\bibinfo  {journal} {arXiv preprint arXiv:2109.12127}\ } (\bibinfo {year} {2021}{\natexlab{a}})}\BibitemShut {NoStop}%
\bibitem [{\citenamefont {J{\'e}rome}\ \emph {et~al.}(1967)\citenamefont {J{\'e}rome}, \citenamefont {Rice},\ and\ \citenamefont {Kohn}}]{jerome1967excitonic}%
  \BibitemOpen
  \bibfield  {author} {\bibinfo {author} {\bibfnamefont {D.}~\bibnamefont {J{\'e}rome}}, \bibinfo {author} {\bibfnamefont {T.}~\bibnamefont {Rice}},\ and\ \bibinfo {author} {\bibfnamefont {W.}~\bibnamefont {Kohn}},\ }\bibfield  {title} {\bibinfo {title} {\textit{Excitonic insulator}},\ }\href@noop {} {\bibfield  {journal} {\bibinfo  {journal} {Phys. Rev.}\ }\textbf {\bibinfo {volume} {158}},\ \bibinfo {pages} {462} (\bibinfo {year} {1967})}\BibitemShut {NoStop}%
\bibitem [{\citenamefont {Kogar}\ \emph {et~al.}(2017)\citenamefont {Kogar}, \citenamefont {Rak}, \citenamefont {Vig}, \citenamefont {Husain}, \citenamefont {Flicker}, \citenamefont {Joe}, \citenamefont {Venema}, \citenamefont {MacDougall}, \citenamefont {Chiang}, \citenamefont {Fradkin} \emph {et~al.}}]{kogar2017signatures}%
  \BibitemOpen
  \bibfield  {author} {\bibinfo {author} {\bibfnamefont {A.}~\bibnamefont {Kogar}}, \bibinfo {author} {\bibfnamefont {M.~S.}\ \bibnamefont {Rak}}, \bibinfo {author} {\bibfnamefont {S.}~\bibnamefont {Vig}}, \bibinfo {author} {\bibfnamefont {A.~A.}\ \bibnamefont {Husain}}, \bibinfo {author} {\bibfnamefont {F.}~\bibnamefont {Flicker}}, \bibinfo {author} {\bibfnamefont {Y.~I.}\ \bibnamefont {Joe}}, \bibinfo {author} {\bibfnamefont {L.}~\bibnamefont {Venema}}, \bibinfo {author} {\bibfnamefont {G.~J.}\ \bibnamefont {MacDougall}}, \bibinfo {author} {\bibfnamefont {T.~C.}\ \bibnamefont {Chiang}}, \bibinfo {author} {\bibfnamefont {E.}~\bibnamefont {Fradkin}}, \emph {et~al.},\ }\bibfield  {title} {\bibinfo {title} {\textit{Signatures of exciton condensation in a transition metal dichalcogenide}},\ }\href@noop {} {\bibfield  {journal} {\bibinfo  {journal} {Science}\ }\textbf {\bibinfo {volume} {358}},\ \bibinfo {pages} {1314} (\bibinfo {year} {2017})}\BibitemShut {NoStop}%
\bibitem [{\citenamefont {Di~Salvo}\ \emph {et~al.}(1986)\citenamefont {Di~Salvo}, \citenamefont {Chen}, \citenamefont {Fleming}, \citenamefont {Waszczak}, \citenamefont {Dunn}, \citenamefont {Sunshine},\ and\ \citenamefont {Ibers}}]{di1986physical}%
  \BibitemOpen
  \bibfield  {author} {\bibinfo {author} {\bibfnamefont {F.}~\bibnamefont {Di~Salvo}}, \bibinfo {author} {\bibfnamefont {C.}~\bibnamefont {Chen}}, \bibinfo {author} {\bibfnamefont {R.}~\bibnamefont {Fleming}}, \bibinfo {author} {\bibfnamefont {J.}~\bibnamefont {Waszczak}}, \bibinfo {author} {\bibfnamefont {R.}~\bibnamefont {Dunn}}, \bibinfo {author} {\bibfnamefont {S.}~\bibnamefont {Sunshine}},\ and\ \bibinfo {author} {\bibfnamefont {J.~A.}\ \bibnamefont {Ibers}},\ }\bibfield  {title} {\bibinfo {title} {\textit{Physical and structural properties of the new layered compounds} {Ta$_2$NiS$_5$} \textit{and} {Ta$_2$NiSe$_5$}},\ }\href@noop {} {\bibfield  {journal} {\bibinfo  {journal} {J. Less-common Met.}\ }\textbf {\bibinfo {volume} {116}},\ \bibinfo {pages} {51} (\bibinfo {year} {1986})}\BibitemShut {NoStop}%
\bibitem [{\citenamefont {Lu}\ \emph {et~al.}(2017)\citenamefont {Lu}, \citenamefont {Kono}, \citenamefont {Larkin}, \citenamefont {Rost}, \citenamefont {Takayama}, \citenamefont {Boris}, \citenamefont {Keimer},\ and\ \citenamefont {Takagi}}]{lu2017zero}%
  \BibitemOpen
  \bibfield  {author} {\bibinfo {author} {\bibfnamefont {Y.}~\bibnamefont {Lu}}, \bibinfo {author} {\bibfnamefont {H.}~\bibnamefont {Kono}}, \bibinfo {author} {\bibfnamefont {T.}~\bibnamefont {Larkin}}, \bibinfo {author} {\bibfnamefont {A.}~\bibnamefont {Rost}}, \bibinfo {author} {\bibfnamefont {T.}~\bibnamefont {Takayama}}, \bibinfo {author} {\bibfnamefont {A.}~\bibnamefont {Boris}}, \bibinfo {author} {\bibfnamefont {B.}~\bibnamefont {Keimer}},\ and\ \bibinfo {author} {\bibfnamefont {H.}~\bibnamefont {Takagi}},\ }\bibfield  {title} {\bibinfo {title} {\textit{Zero-gap semiconductor to excitonic insulator transition in} {Ta$_2$NiSe$_5$}},\ }\href@noop {} {\bibfield  {journal} {\bibinfo  {journal} {Nat. Commun.}\ }\textbf {\bibinfo {volume} {8}},\ \bibinfo {pages} {14408} (\bibinfo {year} {2017})}\BibitemShut {NoStop}%
\bibitem [{\citenamefont {Lee}\ \emph {et~al.}(2019)\citenamefont {Lee}, \citenamefont {Kang}, \citenamefont {Eom}, \citenamefont {Kim}, \citenamefont {Min},\ and\ \citenamefont {Yeom}}]{lee2019strong}%
  \BibitemOpen
  \bibfield  {author} {\bibinfo {author} {\bibfnamefont {J.}~\bibnamefont {Lee}}, \bibinfo {author} {\bibfnamefont {C.-J.}\ \bibnamefont {Kang}}, \bibinfo {author} {\bibfnamefont {M.~J.}\ \bibnamefont {Eom}}, \bibinfo {author} {\bibfnamefont {J.~S.}\ \bibnamefont {Kim}}, \bibinfo {author} {\bibfnamefont {B.~I.}\ \bibnamefont {Min}},\ and\ \bibinfo {author} {\bibfnamefont {H.~W.}\ \bibnamefont {Yeom}},\ }\bibfield  {title} {\bibinfo {title} {\textit{Strong interband interaction in the excitonic insulator phase of} {Ta$_2$NiSe$_5$}},\ }\href@noop {} {\bibfield  {journal} {\bibinfo  {journal} {Phys. Rev. B}\ }\textbf {\bibinfo {volume} {99}},\ \bibinfo {pages} {075408} (\bibinfo {year} {2019})}\BibitemShut {NoStop}%
\bibitem [{\citenamefont {Wakisaka}\ \emph {et~al.}(2009)\citenamefont {Wakisaka}, \citenamefont {Sudayama}, \citenamefont {Takubo}, \citenamefont {Mizokawa}, \citenamefont {Arita}, \citenamefont {Namatame}, \citenamefont {Taniguchi}, \citenamefont {Katayama}, \citenamefont {Nohara},\ and\ \citenamefont {Takagi}}]{wakisaka2009excitonic}%
  \BibitemOpen
  \bibfield  {author} {\bibinfo {author} {\bibfnamefont {Y.}~\bibnamefont {Wakisaka}}, \bibinfo {author} {\bibfnamefont {T.}~\bibnamefont {Sudayama}}, \bibinfo {author} {\bibfnamefont {K.}~\bibnamefont {Takubo}}, \bibinfo {author} {\bibfnamefont {T.}~\bibnamefont {Mizokawa}}, \bibinfo {author} {\bibfnamefont {M.}~\bibnamefont {Arita}}, \bibinfo {author} {\bibfnamefont {H.}~\bibnamefont {Namatame}}, \bibinfo {author} {\bibfnamefont {M.}~\bibnamefont {Taniguchi}}, \bibinfo {author} {\bibfnamefont {N.}~\bibnamefont {Katayama}}, \bibinfo {author} {\bibfnamefont {M.}~\bibnamefont {Nohara}},\ and\ \bibinfo {author} {\bibfnamefont {H.}~\bibnamefont {Takagi}},\ }\bibfield  {title} {\bibinfo {title} {\textit{Excitonic insulator state in} {Ta$_2$NiSe$_5$} \textit{probed by photoemission spectroscopy}},\ }\href@noop {} {\bibfield  {journal} {\bibinfo  {journal} {Phys. Rev. Lett.}\ }\textbf {\bibinfo {volume} {103}},\ \bibinfo {pages} {026402} (\bibinfo {year} {2009})}\BibitemShut {NoStop}%
\bibitem [{\citenamefont {Wakisaka}\ \emph {et~al.}(2012)\citenamefont {Wakisaka}, \citenamefont {Sudayama}, \citenamefont {Takubo}, \citenamefont {Mizokawa}, \citenamefont {Saini}, \citenamefont {Arita}, \citenamefont {Namatame}, \citenamefont {Taniguchi}, \citenamefont {Katayama}, \citenamefont {Nohara} \emph {et~al.}}]{wakisaka2012photoemission}%
  \BibitemOpen
  \bibfield  {author} {\bibinfo {author} {\bibfnamefont {Y.}~\bibnamefont {Wakisaka}}, \bibinfo {author} {\bibfnamefont {T.}~\bibnamefont {Sudayama}}, \bibinfo {author} {\bibfnamefont {K.}~\bibnamefont {Takubo}}, \bibinfo {author} {\bibfnamefont {T.}~\bibnamefont {Mizokawa}}, \bibinfo {author} {\bibfnamefont {N.}~\bibnamefont {Saini}}, \bibinfo {author} {\bibfnamefont {M.}~\bibnamefont {Arita}}, \bibinfo {author} {\bibfnamefont {H.}~\bibnamefont {Namatame}}, \bibinfo {author} {\bibfnamefont {M.}~\bibnamefont {Taniguchi}}, \bibinfo {author} {\bibfnamefont {N.}~\bibnamefont {Katayama}}, \bibinfo {author} {\bibfnamefont {M.}~\bibnamefont {Nohara}}, \emph {et~al.},\ }\bibfield  {title} {\bibinfo {title} {\textit{Photoemission spectroscopy of} {Ta$_2$NiSe$_5$}},\ }\href@noop {} {\bibfield  {journal} {\bibinfo  {journal} {J. Supercond. Nov. Magn.}\ }\textbf {\bibinfo {volume} {25}},\ \bibinfo {pages} {1231} (\bibinfo {year} {2012})}\BibitemShut {NoStop}%
\bibitem [{\citenamefont {Seki}\ \emph {et~al.}(2014)\citenamefont {Seki}, \citenamefont {Wakisaka}, \citenamefont {Kaneko}, \citenamefont {Toriyama}, \citenamefont {Konishi}, \citenamefont {Sudayama}, \citenamefont {Saini}, \citenamefont {Arita}, \citenamefont {Namatame}, \citenamefont {Taniguchi} \emph {et~al.}}]{seki2014excitonic}%
  \BibitemOpen
  \bibfield  {author} {\bibinfo {author} {\bibfnamefont {K.}~\bibnamefont {Seki}}, \bibinfo {author} {\bibfnamefont {Y.}~\bibnamefont {Wakisaka}}, \bibinfo {author} {\bibfnamefont {T.}~\bibnamefont {Kaneko}}, \bibinfo {author} {\bibfnamefont {T.}~\bibnamefont {Toriyama}}, \bibinfo {author} {\bibfnamefont {T.}~\bibnamefont {Konishi}}, \bibinfo {author} {\bibfnamefont {T.}~\bibnamefont {Sudayama}}, \bibinfo {author} {\bibfnamefont {N.}~\bibnamefont {Saini}}, \bibinfo {author} {\bibfnamefont {M.}~\bibnamefont {Arita}}, \bibinfo {author} {\bibfnamefont {H.}~\bibnamefont {Namatame}}, \bibinfo {author} {\bibfnamefont {M.}~\bibnamefont {Taniguchi}}, \emph {et~al.},\ }\bibfield  {title} {\bibinfo {title} {\textit{Excitonic bose-einstein condensation in} {Ta$_2$NiSe$_5$} \textit{above room temperature}},\ }\href@noop {} {\bibfield  {journal} {\bibinfo  {journal} {Phys. Rev. B}\ }\textbf {\bibinfo {volume} {90}},\ \bibinfo {pages} {155116} (\bibinfo {year} {2014})}\BibitemShut {NoStop}%
\bibitem [{\citenamefont {Chen}\ \emph {et~al.}(2022)\citenamefont {Chen}, \citenamefont {Chen}, \citenamefont {Tang}, \citenamefont {Li}, \citenamefont {Wang}, \citenamefont {Ding}, \citenamefont {Kang}, \citenamefont {Jozwiak}, \citenamefont {Bostwick}, \citenamefont {Rotenberg}, \citenamefont {Hashimoto} \emph {et~al.}}]{chen2022lattice}%
  \BibitemOpen
  \bibfield  {author} {\bibinfo {author} {\bibfnamefont {C.}~\bibnamefont {Chen}}, \bibinfo {author} {\bibfnamefont {X.}~\bibnamefont {Chen}}, \bibinfo {author} {\bibfnamefont {W.}~\bibnamefont {Tang}}, \bibinfo {author} {\bibfnamefont {Z.}~\bibnamefont {Li}}, \bibinfo {author} {\bibfnamefont {S.}~\bibnamefont {Wang}}, \bibinfo {author} {\bibfnamefont {S.}~\bibnamefont {Ding}}, \bibinfo {author} {\bibfnamefont {Z.}~\bibnamefont {Kang}}, \bibinfo {author} {\bibfnamefont {C.}~\bibnamefont {Jozwiak}}, \bibinfo {author} {\bibfnamefont {A.}~\bibnamefont {Bostwick}}, \bibinfo {author} {\bibfnamefont {E.}~\bibnamefont {Rotenberg}}, \bibinfo {author} {\bibfnamefont {M.}~\bibnamefont {Hashimoto}}, \emph {et~al.},\ }\bibfield  {title} {\bibinfo {title} {\textit{Role of electron-phonon coupling in excitonic insulator candidate} {Ta$_2$NiSe$_5$}},\ }\href@noop {} {\bibfield  {journal} {\bibinfo  {journal} {arXiv preprint arXiv:2203.06817v2}\ } (\bibinfo {year} {2022})}\BibitemShut {NoStop}%
\bibitem [{\citenamefont {Mazza}\ \emph {et~al.}(2020)\citenamefont {Mazza}, \citenamefont {R{\"o}sner}, \citenamefont {Windg{\"a}tter}, \citenamefont {Latini}, \citenamefont {H{\"u}bener}, \citenamefont {Millis}, \citenamefont {Rubio},\ and\ \citenamefont {Georges}}]{mazza2020nature}%
  \BibitemOpen
  \bibfield  {author} {\bibinfo {author} {\bibfnamefont {G.}~\bibnamefont {Mazza}}, \bibinfo {author} {\bibfnamefont {M.}~\bibnamefont {R{\"o}sner}}, \bibinfo {author} {\bibfnamefont {L.}~\bibnamefont {Windg{\"a}tter}}, \bibinfo {author} {\bibfnamefont {S.}~\bibnamefont {Latini}}, \bibinfo {author} {\bibfnamefont {H.}~\bibnamefont {H{\"u}bener}}, \bibinfo {author} {\bibfnamefont {A.~J.}\ \bibnamefont {Millis}}, \bibinfo {author} {\bibfnamefont {A.}~\bibnamefont {Rubio}},\ and\ \bibinfo {author} {\bibfnamefont {A.}~\bibnamefont {Georges}},\ }\bibfield  {title} {\bibinfo {title} {\textit{Nature of symmetry breaking at the excitonic insulator transition:} {Ta$_2$NiSe$_5$}},\ }\href@noop {} {\bibfield  {journal} {\bibinfo  {journal} {Phys. Rev. Lett.}\ }\textbf {\bibinfo {volume} {124}},\ \bibinfo {pages} {197601} (\bibinfo {year} {2020})}\BibitemShut {NoStop}%
\bibitem [{\citenamefont {Baldini}\ \emph {et~al.}(2023)\citenamefont {Baldini}, \citenamefont {Zong}, \citenamefont {Choi}, \citenamefont {Lee}, \citenamefont {Michael}, \citenamefont {Windgaetter}, \citenamefont {Mazin}, \citenamefont {Latini}, \citenamefont {Azoury}, \citenamefont {Lv} \emph {et~al.}}]{baldini2023spontaneous}%
  \BibitemOpen
  \bibfield  {author} {\bibinfo {author} {\bibfnamefont {E.}~\bibnamefont {Baldini}}, \bibinfo {author} {\bibfnamefont {A.}~\bibnamefont {Zong}}, \bibinfo {author} {\bibfnamefont {D.}~\bibnamefont {Choi}}, \bibinfo {author} {\bibfnamefont {C.}~\bibnamefont {Lee}}, \bibinfo {author} {\bibfnamefont {M.~H.}\ \bibnamefont {Michael}}, \bibinfo {author} {\bibfnamefont {L.}~\bibnamefont {Windgaetter}}, \bibinfo {author} {\bibfnamefont {I.~I.}\ \bibnamefont {Mazin}}, \bibinfo {author} {\bibfnamefont {S.}~\bibnamefont {Latini}}, \bibinfo {author} {\bibfnamefont {D.}~\bibnamefont {Azoury}}, \bibinfo {author} {\bibfnamefont {B.}~\bibnamefont {Lv}}, \emph {et~al.},\ }\bibfield  {title} {\bibinfo {title} {\textit{The spontaneous symmetry breaking in} {Ta$_2$NiSe$_5$} \textit{is structural in nature}},\ }\href@noop {} {\bibfield  {journal} {\bibinfo  {journal} {Proc. Natl. Acad. Sci.}\ }\textbf {\bibinfo {volume} {120}},\ \bibinfo {pages} {e2221688120} (\bibinfo {year} {2023})}\BibitemShut {NoStop}%
\bibitem [{\citenamefont {Watson}\ \emph {et~al.}(2020)\citenamefont {Watson}, \citenamefont {Markovi{\'c}}, \citenamefont {Morales}, \citenamefont {Le~F{\`e}vre}, \citenamefont {Merz}, \citenamefont {Haghighirad},\ and\ \citenamefont {King}}]{watson2020band}%
  \BibitemOpen
  \bibfield  {author} {\bibinfo {author} {\bibfnamefont {M.~D.}\ \bibnamefont {Watson}}, \bibinfo {author} {\bibfnamefont {I.}~\bibnamefont {Markovi{\'c}}}, \bibinfo {author} {\bibfnamefont {E.~A.}\ \bibnamefont {Morales}}, \bibinfo {author} {\bibfnamefont {P.}~\bibnamefont {Le~F{\`e}vre}}, \bibinfo {author} {\bibfnamefont {M.}~\bibnamefont {Merz}}, \bibinfo {author} {\bibfnamefont {A.~A.}\ \bibnamefont {Haghighirad}},\ and\ \bibinfo {author} {\bibfnamefont {P.~D.}\ \bibnamefont {King}},\ }\bibfield  {title} {\bibinfo {title} {\textit{Band hybridization at the semimetal-semiconductor transition of} {Ta$_2$NiSe$_5$} \textit{enabled by mirror-symmetry breaking}},\ }\href@noop {} {\bibfield  {journal} {\bibinfo  {journal} {Phys. Rev. Res.}\ }\textbf {\bibinfo {volume} {2}},\ \bibinfo {pages} {013236} (\bibinfo {year} {2020})}\BibitemShut {NoStop}%
\bibitem [{\citenamefont {Bretscher}\ \emph {et~al.}(2021{\natexlab{a}})\citenamefont {Bretscher}, \citenamefont {Andrich}, \citenamefont {Telang}, \citenamefont {Singh}, \citenamefont {Harnagea}, \citenamefont {Sood},\ and\ \citenamefont {Rao}}]{bretscher2021ultrafast}%
  \BibitemOpen
  \bibfield  {author} {\bibinfo {author} {\bibfnamefont {H.~M.}\ \bibnamefont {Bretscher}}, \bibinfo {author} {\bibfnamefont {P.}~\bibnamefont {Andrich}}, \bibinfo {author} {\bibfnamefont {P.}~\bibnamefont {Telang}}, \bibinfo {author} {\bibfnamefont {A.}~\bibnamefont {Singh}}, \bibinfo {author} {\bibfnamefont {L.}~\bibnamefont {Harnagea}}, \bibinfo {author} {\bibfnamefont {A.~K.}\ \bibnamefont {Sood}},\ and\ \bibinfo {author} {\bibfnamefont {A.}~\bibnamefont {Rao}},\ }\bibfield  {title} {\bibinfo {title} {\textit{Ultrafast melting and recovery of collective order in the excitonic insulator} {Ta$_2$NiSe$_5$}},\ }\href@noop {} {\bibfield  {journal} {\bibinfo  {journal} {Nat. Commun.}\ }\textbf {\bibinfo {volume} {12}},\ \bibinfo {pages} {1699} (\bibinfo {year} {2021}{\natexlab{a}})}\BibitemShut {NoStop}%
\bibitem [{\citenamefont {Yan}\ \emph {et~al.}(2019)\citenamefont {Yan}, \citenamefont {Xiao}, \citenamefont {Luo}, \citenamefont {Lv}, \citenamefont {Zhang}, \citenamefont {Sun}, \citenamefont {Tong}, \citenamefont {Lu}, \citenamefont {Song}, \citenamefont {Zhu} \emph {et~al.}}]{yan2019strong}%
  \BibitemOpen
  \bibfield  {author} {\bibinfo {author} {\bibfnamefont {J.}~\bibnamefont {Yan}}, \bibinfo {author} {\bibfnamefont {R.}~\bibnamefont {Xiao}}, \bibinfo {author} {\bibfnamefont {X.}~\bibnamefont {Luo}}, \bibinfo {author} {\bibfnamefont {H.}~\bibnamefont {Lv}}, \bibinfo {author} {\bibfnamefont {R.}~\bibnamefont {Zhang}}, \bibinfo {author} {\bibfnamefont {Y.}~\bibnamefont {Sun}}, \bibinfo {author} {\bibfnamefont {P.}~\bibnamefont {Tong}}, \bibinfo {author} {\bibfnamefont {W.}~\bibnamefont {Lu}}, \bibinfo {author} {\bibfnamefont {W.}~\bibnamefont {Song}}, \bibinfo {author} {\bibfnamefont {X.}~\bibnamefont {Zhu}}, \emph {et~al.},\ }\bibfield  {title} {\bibinfo {title} {\textit{Strong electron-phonon coupling in the excitonic insulator} {Ta$_2$NiSe$_5$}},\ }\href@noop {} {\bibfield  {journal} {\bibinfo  {journal} {Inorg. Chem.}\ }\textbf {\bibinfo {volume} {58}},\ \bibinfo {pages} {9036} (\bibinfo {year} {2019})}\BibitemShut {NoStop}%
\bibitem [{\citenamefont {Volkov}\ \emph {et~al.}(2021{\natexlab{a}})\citenamefont {Volkov}, \citenamefont {Ye}, \citenamefont {Lohani}, \citenamefont {Feldman}, \citenamefont {Kanigel},\ and\ \citenamefont {Blumberg}}]{volkov2021critical}%
  \BibitemOpen
  \bibfield  {author} {\bibinfo {author} {\bibfnamefont {P.~A.}\ \bibnamefont {Volkov}}, \bibinfo {author} {\bibfnamefont {M.}~\bibnamefont {Ye}}, \bibinfo {author} {\bibfnamefont {H.}~\bibnamefont {Lohani}}, \bibinfo {author} {\bibfnamefont {I.}~\bibnamefont {Feldman}}, \bibinfo {author} {\bibfnamefont {A.}~\bibnamefont {Kanigel}},\ and\ \bibinfo {author} {\bibfnamefont {G.}~\bibnamefont {Blumberg}},\ }\bibfield  {title} {\bibinfo {title} {\textit{Critical charge fluctuations and emergent coherence in a strongly correlated excitonic insulator}},\ }\href@noop {} {\bibfield  {journal} {\bibinfo  {journal} {npj Quantum Mater.}\ }\textbf {\bibinfo {volume} {6}},\ \bibinfo {pages} {52} (\bibinfo {year} {2021}{\natexlab{a}})}\BibitemShut {NoStop}%
\bibitem [{\citenamefont {Volkov}\ \emph {et~al.}(2021{\natexlab{b}})\citenamefont {Volkov}, \citenamefont {Ye}, \citenamefont {Lohani}, \citenamefont {Feldman}, \citenamefont {Kanigel},\ and\ \citenamefont {Blumberg}}]{volkov2021failed}%
  \BibitemOpen
  \bibfield  {author} {\bibinfo {author} {\bibfnamefont {P.~A.}\ \bibnamefont {Volkov}}, \bibinfo {author} {\bibfnamefont {M.}~\bibnamefont {Ye}}, \bibinfo {author} {\bibfnamefont {H.}~\bibnamefont {Lohani}}, \bibinfo {author} {\bibfnamefont {I.}~\bibnamefont {Feldman}}, \bibinfo {author} {\bibfnamefont {A.}~\bibnamefont {Kanigel}},\ and\ \bibinfo {author} {\bibfnamefont {G.}~\bibnamefont {Blumberg}},\ }\bibfield  {title} {\bibinfo {title} {\textit{Failed excitonic quantum phase transition in} {Ta$_2$Ni(Se$_{1-x}$S$_x$)$_5$}},\ }\href@noop {} {\bibfield  {journal} {\bibinfo  {journal} {Phys. Rev. B}\ }\textbf {\bibinfo {volume} {104}},\ \bibinfo {pages} {L241103} (\bibinfo {year} {2021}{\natexlab{b}})}\BibitemShut {NoStop}%
\bibitem [{\citenamefont {Kim}\ \emph {et~al.}(2021{\natexlab{b}})\citenamefont {Kim}, \citenamefont {Kim}, \citenamefont {Kim}, \citenamefont {Kwon}, \citenamefont {Kim},\ and\ \citenamefont {Kim}}]{kim2021direct}%
  \BibitemOpen
  \bibfield  {author} {\bibinfo {author} {\bibfnamefont {K.}~\bibnamefont {Kim}}, \bibinfo {author} {\bibfnamefont {H.}~\bibnamefont {Kim}}, \bibinfo {author} {\bibfnamefont {J.}~\bibnamefont {Kim}}, \bibinfo {author} {\bibfnamefont {C.}~\bibnamefont {Kwon}}, \bibinfo {author} {\bibfnamefont {J.~S.}\ \bibnamefont {Kim}},\ and\ \bibinfo {author} {\bibfnamefont {B.}~\bibnamefont {Kim}},\ }\bibfield  {title} {\bibinfo {title} {\textit{Direct observation of excitonic instability in} {Ta$_2$NiSe$_5$}},\ }\href@noop {} {\bibfield  {journal} {\bibinfo  {journal} {Nat. Commun.}\ }\textbf {\bibinfo {volume} {12}},\ \bibinfo {pages} {1969} (\bibinfo {year} {2021}{\natexlab{b}})}\BibitemShut {NoStop}%
\bibitem [{\citenamefont {Larkin}\ \emph {et~al.}(2017)\citenamefont {Larkin}, \citenamefont {Yaresko}, \citenamefont {Pr{\"o}pper}, \citenamefont {Kikoin}, \citenamefont {Lu}, \citenamefont {Takayama}, \citenamefont {Mathis}, \citenamefont {Rost}, \citenamefont {Takagi}, \citenamefont {Keimer} \emph {et~al.}}]{larkin2017giant}%
  \BibitemOpen
  \bibfield  {author} {\bibinfo {author} {\bibfnamefont {T.}~\bibnamefont {Larkin}}, \bibinfo {author} {\bibfnamefont {A.}~\bibnamefont {Yaresko}}, \bibinfo {author} {\bibfnamefont {D.}~\bibnamefont {Pr{\"o}pper}}, \bibinfo {author} {\bibfnamefont {K.}~\bibnamefont {Kikoin}}, \bibinfo {author} {\bibfnamefont {Y.}~\bibnamefont {Lu}}, \bibinfo {author} {\bibfnamefont {T.}~\bibnamefont {Takayama}}, \bibinfo {author} {\bibfnamefont {Y.-L.}\ \bibnamefont {Mathis}}, \bibinfo {author} {\bibfnamefont {A.}~\bibnamefont {Rost}}, \bibinfo {author} {\bibfnamefont {H.}~\bibnamefont {Takagi}}, \bibinfo {author} {\bibfnamefont {B.}~\bibnamefont {Keimer}}, \emph {et~al.},\ }\bibfield  {title} {\bibinfo {title} {\textit{Giant exciton Fano resonance in quasi-one-dimensional} {Ta$_2$NiSe$_5$}},\ }\href@noop {} {\bibfield  {journal} {\bibinfo  {journal} {Phys. Rev. B}\ }\textbf {\bibinfo {volume} {95}},\ \bibinfo {pages} {195144} (\bibinfo {year} {2017})}\BibitemShut {NoStop}%
\bibitem [{\citenamefont {Larkin}\ \emph {et~al.}(2018)\citenamefont {Larkin}, \citenamefont {Dawson}, \citenamefont {H{\"o}ppner}, \citenamefont {Takayama}, \citenamefont {Isobe}, \citenamefont {Mathis}, \citenamefont {Takagi}, \citenamefont {Keimer},\ and\ \citenamefont {Boris}}]{larkin2018infrared}%
  \BibitemOpen
  \bibfield  {author} {\bibinfo {author} {\bibfnamefont {T.}~\bibnamefont {Larkin}}, \bibinfo {author} {\bibfnamefont {R.}~\bibnamefont {Dawson}}, \bibinfo {author} {\bibfnamefont {M.}~\bibnamefont {H{\"o}ppner}}, \bibinfo {author} {\bibfnamefont {T.}~\bibnamefont {Takayama}}, \bibinfo {author} {\bibfnamefont {M.}~\bibnamefont {Isobe}}, \bibinfo {author} {\bibfnamefont {Y.-L.}\ \bibnamefont {Mathis}}, \bibinfo {author} {\bibfnamefont {H.}~\bibnamefont {Takagi}}, \bibinfo {author} {\bibfnamefont {B.}~\bibnamefont {Keimer}},\ and\ \bibinfo {author} {\bibfnamefont {A.}~\bibnamefont {Boris}},\ }\bibfield  {title} {\bibinfo {title} {\textit{Infrared phonon spectra of quasi-one-dimensional} {Ta$_2$NiSe$_5$} \textit{and} {Ta$_2$NiS$_5$}},\ }\href@noop {} {\bibfield  {journal} {\bibinfo  {journal} {Phys. Rev. B}\ }\textbf {\bibinfo {volume} {98}},\ \bibinfo {pages} {125113} (\bibinfo {year} {2018})}\BibitemShut {NoStop}%
\bibitem [{\citenamefont {Werdehausen}\ \emph {et~al.}(2018)\citenamefont {Werdehausen}, \citenamefont {Takayama}, \citenamefont {H{\"o}ppner}, \citenamefont {Albrecht}, \citenamefont {Rost}, \citenamefont {Lu}, \citenamefont {Manske}, \citenamefont {Takagi},\ and\ \citenamefont {Kaiser}}]{werdehausen2018coherent}%
  \BibitemOpen
  \bibfield  {author} {\bibinfo {author} {\bibfnamefont {D.}~\bibnamefont {Werdehausen}}, \bibinfo {author} {\bibfnamefont {T.}~\bibnamefont {Takayama}}, \bibinfo {author} {\bibfnamefont {M.}~\bibnamefont {H{\"o}ppner}}, \bibinfo {author} {\bibfnamefont {G.}~\bibnamefont {Albrecht}}, \bibinfo {author} {\bibfnamefont {A.~W.}\ \bibnamefont {Rost}}, \bibinfo {author} {\bibfnamefont {Y.}~\bibnamefont {Lu}}, \bibinfo {author} {\bibfnamefont {D.}~\bibnamefont {Manske}}, \bibinfo {author} {\bibfnamefont {H.}~\bibnamefont {Takagi}},\ and\ \bibinfo {author} {\bibfnamefont {S.}~\bibnamefont {Kaiser}},\ }\bibfield  {title} {\bibinfo {title} {\textit{Coherent order parameter oscillations in the ground state of the excitonic insulator} {Ta$_2$NiSe$_5$}},\ }\href@noop {} {\bibfield  {journal} {\bibinfo  {journal} {Sci. Adv.}\ }\textbf {\bibinfo {volume} {4}},\ \bibinfo {pages} {eaap8652} (\bibinfo {year} {2018})}\BibitemShut {NoStop}%
\bibitem [{\citenamefont {Bretscher}\ \emph {et~al.}(2021{\natexlab{b}})\citenamefont {Bretscher}, \citenamefont {Andrich}, \citenamefont {Murakami}, \citenamefont {Gole{\v{z}}}, \citenamefont {Remez}, \citenamefont {Telang}, \citenamefont {Singh}, \citenamefont {Harnagea}, \citenamefont {Cooper}, \citenamefont {Millis} \emph {et~al.}}]{bretscher2021imaging}%
  \BibitemOpen
  \bibfield  {author} {\bibinfo {author} {\bibfnamefont {H.~M.}\ \bibnamefont {Bretscher}}, \bibinfo {author} {\bibfnamefont {P.}~\bibnamefont {Andrich}}, \bibinfo {author} {\bibfnamefont {Y.}~\bibnamefont {Murakami}}, \bibinfo {author} {\bibfnamefont {D.}~\bibnamefont {Gole{\v{z}}}}, \bibinfo {author} {\bibfnamefont {B.}~\bibnamefont {Remez}}, \bibinfo {author} {\bibfnamefont {P.}~\bibnamefont {Telang}}, \bibinfo {author} {\bibfnamefont {A.}~\bibnamefont {Singh}}, \bibinfo {author} {\bibfnamefont {L.}~\bibnamefont {Harnagea}}, \bibinfo {author} {\bibfnamefont {N.~R.}\ \bibnamefont {Cooper}}, \bibinfo {author} {\bibfnamefont {A.~J.}\ \bibnamefont {Millis}}, \emph {et~al.},\ }\bibfield  {title} {\bibinfo {title} {\textit{Imaging the coherent propagation of collective modes in the excitonic insulator} {Ta$_2$NiSe$_5$} \textit{at room temperature}},\ }\href@noop {} {\bibfield  {journal} {\bibinfo  {journal} {Sci. Adv.}\ }\textbf {\bibinfo {volume} {7}},\ \bibinfo {pages} {eabd6147} (\bibinfo {year}
  {2021}{\natexlab{b}})}\BibitemShut {NoStop}%
\bibitem [{\citenamefont {Saha}\ \emph {et~al.}(2022)\citenamefont {Saha}, \citenamefont {Petaccia}, \citenamefont {Ressel}, \citenamefont {Ribi\ifmmode~\check{c}\else \v{c}\fi{}}, \citenamefont {Di~Santo}, \citenamefont {Zhao},\ and\ \citenamefont {De~Ninno}}]{Tanusree2022electronic}%
  \BibitemOpen
  \bibfield  {author} {\bibinfo {author} {\bibfnamefont {T.}~\bibnamefont {Saha}}, \bibinfo {author} {\bibfnamefont {L.}~\bibnamefont {Petaccia}}, \bibinfo {author} {\bibfnamefont {B.}~\bibnamefont {Ressel}}, \bibinfo {author} {\bibfnamefont {P.~c. v.~R.}\ \bibnamefont {Ribi\ifmmode~\check{c}\else \v{c}\fi{}}}, \bibinfo {author} {\bibfnamefont {G.}~\bibnamefont {Di~Santo}}, \bibinfo {author} {\bibfnamefont {W.}~\bibnamefont {Zhao}},\ and\ \bibinfo {author} {\bibfnamefont {G.}~\bibnamefont {De~Ninno}},\ }\bibfield  {title} {\bibinfo {title} {\textit{Electronic band structure in pristine and sulfur-doped} {${\mathrm{Ta}}_{2}{\mathrm{NiSe}}_{5}$}},\ }\href@noop {} {\bibfield  {journal} {\bibinfo  {journal} {Phys. Rev. B}\ }\textbf {\bibinfo {volume} {106}},\ \bibinfo {pages} {075148} (\bibinfo {year} {2022})}\BibitemShut {NoStop}%
\bibitem [{\citenamefont {Sugimoto}\ \emph {et~al.}(2018)\citenamefont {Sugimoto}, \citenamefont {Nishimoto}, \citenamefont {Kaneko},\ and\ \citenamefont {Ohta}}]{sugimoto2018strong}%
  \BibitemOpen
  \bibfield  {author} {\bibinfo {author} {\bibfnamefont {K.}~\bibnamefont {Sugimoto}}, \bibinfo {author} {\bibfnamefont {S.}~\bibnamefont {Nishimoto}}, \bibinfo {author} {\bibfnamefont {T.}~\bibnamefont {Kaneko}},\ and\ \bibinfo {author} {\bibfnamefont {Y.}~\bibnamefont {Ohta}},\ }\bibfield  {title} {\bibinfo {title} {\textit{Strong coupling nature of the excitonic insulator state in} {Ta$_2$NiSe$_5$}},\ }\href@noop {} {\bibfield  {journal} {\bibinfo  {journal} {Phys. Rev. Lett.}\ }\textbf {\bibinfo {volume} {120}},\ \bibinfo {pages} {247602} (\bibinfo {year} {2018})}\BibitemShut {NoStop}%
\bibitem [{\citenamefont {Zhu}\ \emph {et~al.}(2015)\citenamefont {Zhu}, \citenamefont {Cao}, \citenamefont {Zhang}, \citenamefont {Plummer},\ and\ \citenamefont {Guo}}]{zhu2015classification}%
  \BibitemOpen
  \bibfield  {author} {\bibinfo {author} {\bibfnamefont {X.}~\bibnamefont {Zhu}}, \bibinfo {author} {\bibfnamefont {Y.}~\bibnamefont {Cao}}, \bibinfo {author} {\bibfnamefont {J.}~\bibnamefont {Zhang}}, \bibinfo {author} {\bibfnamefont {E.}~\bibnamefont {Plummer}},\ and\ \bibinfo {author} {\bibfnamefont {J.}~\bibnamefont {Guo}},\ }\bibfield  {title} {\bibinfo {title} {\textit{Classification of charge density waves based on their nature}},\ }\href@noop {} {\bibfield  {journal} {\bibinfo  {journal} {Proc. Natl. Acad. Sci.}\ }\textbf {\bibinfo {volume} {112}},\ \bibinfo {pages} {2367} (\bibinfo {year} {2015})}\BibitemShut {NoStop}%
\bibitem [{\citenamefont {Fukutani}\ \emph {et~al.}(2021)\citenamefont {Fukutani}, \citenamefont {Stania}, \citenamefont {Il~Kwon}, \citenamefont {Kim}, \citenamefont {Kong}, \citenamefont {Kim},\ and\ \citenamefont {Yeom}}]{fukutani2021detecting}%
  \BibitemOpen
  \bibfield  {author} {\bibinfo {author} {\bibfnamefont {K.}~\bibnamefont {Fukutani}}, \bibinfo {author} {\bibfnamefont {R.}~\bibnamefont {Stania}}, \bibinfo {author} {\bibfnamefont {C.}~\bibnamefont {Il~Kwon}}, \bibinfo {author} {\bibfnamefont {J.~S.}\ \bibnamefont {Kim}}, \bibinfo {author} {\bibfnamefont {K.~J.}\ \bibnamefont {Kong}}, \bibinfo {author} {\bibfnamefont {J.}~\bibnamefont {Kim}},\ and\ \bibinfo {author} {\bibfnamefont {H.~W.}\ \bibnamefont {Yeom}},\ }\bibfield  {title} {\bibinfo {title} {\textit{Detecting photoelectrons from spontaneously formed excitons}},\ }\href@noop {} {\bibfield  {journal} {\bibinfo  {journal} {Nat. Phys.}\ }\textbf {\bibinfo {volume} {17}},\ \bibinfo {pages} {1024} (\bibinfo {year} {2021})}\BibitemShut {NoStop}%
\bibitem [{\citenamefont {He}\ \emph {et~al.}(2018)\citenamefont {He}, \citenamefont {Hashimoto}, \citenamefont {Song}, \citenamefont {Chen}, \citenamefont {He}, \citenamefont {Vishik}, \citenamefont {Moritz}, \citenamefont {Lee}, \citenamefont {Nagaosa}, \citenamefont {Zaanen} \emph {et~al.}}]{he2018rapid}%
  \BibitemOpen
  \bibfield  {author} {\bibinfo {author} {\bibfnamefont {Y.}~\bibnamefont {He}}, \bibinfo {author} {\bibfnamefont {M.}~\bibnamefont {Hashimoto}}, \bibinfo {author} {\bibfnamefont {D.}~\bibnamefont {Song}}, \bibinfo {author} {\bibfnamefont {S.-D.}\ \bibnamefont {Chen}}, \bibinfo {author} {\bibfnamefont {J.}~\bibnamefont {He}}, \bibinfo {author} {\bibfnamefont {I.}~\bibnamefont {Vishik}}, \bibinfo {author} {\bibfnamefont {B.}~\bibnamefont {Moritz}}, \bibinfo {author} {\bibfnamefont {D.-H.}\ \bibnamefont {Lee}}, \bibinfo {author} {\bibfnamefont {N.}~\bibnamefont {Nagaosa}}, \bibinfo {author} {\bibfnamefont {J.}~\bibnamefont {Zaanen}}, \emph {et~al.},\ }\bibfield  {title} {\bibinfo {title} {\textit{Rapid change of superconductivity and electron-phonon coupling through critical doping in} {Bi-2212}},\ }\href@noop {} {\bibfield  {journal} {\bibinfo  {journal} {Science}\ }\textbf {\bibinfo {volume} {362}},\ \bibinfo {pages} {62} (\bibinfo {year} {2018})}\BibitemShut {NoStop}%
\bibitem [{\citenamefont {Chen}\ \emph {et~al.}(2019)\citenamefont {Chen}, \citenamefont {Hashimoto}, \citenamefont {He}, \citenamefont {Song}, \citenamefont {Xu}, \citenamefont {He}, \citenamefont {Devereaux}, \citenamefont {Eisaki}, \citenamefont {Lu}, \citenamefont {Zaanen} \emph {et~al.}}]{chen2019incoherent}%
  \BibitemOpen
  \bibfield  {author} {\bibinfo {author} {\bibfnamefont {S.-D.}\ \bibnamefont {Chen}}, \bibinfo {author} {\bibfnamefont {M.}~\bibnamefont {Hashimoto}}, \bibinfo {author} {\bibfnamefont {Y.}~\bibnamefont {He}}, \bibinfo {author} {\bibfnamefont {D.}~\bibnamefont {Song}}, \bibinfo {author} {\bibfnamefont {K.-J.}\ \bibnamefont {Xu}}, \bibinfo {author} {\bibfnamefont {J.-F.}\ \bibnamefont {He}}, \bibinfo {author} {\bibfnamefont {T.~P.}\ \bibnamefont {Devereaux}}, \bibinfo {author} {\bibfnamefont {H.}~\bibnamefont {Eisaki}}, \bibinfo {author} {\bibfnamefont {D.-H.}\ \bibnamefont {Lu}}, \bibinfo {author} {\bibfnamefont {J.}~\bibnamefont {Zaanen}}, \emph {et~al.},\ }\bibfield  {title} {\bibinfo {title} {\textit{Incoherent strange metal sharply bounded by a critical doping in} {Bi2212}},\ }\href@noop {} {\bibfield  {journal} {\bibinfo  {journal} {Science}\ }\textbf {\bibinfo {volume} {366}},\ \bibinfo {pages} {1099} (\bibinfo {year} {2019})}\BibitemShut {NoStop}%
\bibitem [{\citenamefont {Phan}\ \emph {et~al.}(2010)\citenamefont {Phan}, \citenamefont {Becker},\ and\ \citenamefont {Fehske}}]{hogler2010spectral}%
  \BibitemOpen
  \bibfield  {author} {\bibinfo {author} {\bibfnamefont {V.-N.}\ \bibnamefont {Phan}}, \bibinfo {author} {\bibfnamefont {K.~W.}\ \bibnamefont {Becker}},\ and\ \bibinfo {author} {\bibfnamefont {H.}~\bibnamefont {Fehske}},\ }\bibfield  {title} {\bibinfo {title} {\textit{Spectral signatures of the BCS-BEC crossover in the excitonic insulator phase of the extended Falicov-Kimball model}},\ }\href@noop {} {\bibfield  {journal} {\bibinfo  {journal} {Phys. Rev. B}\ }\textbf {\bibinfo {volume} {81}},\ \bibinfo {pages} {205117} (\bibinfo {year} {2010})}\BibitemShut {NoStop}%
\bibitem [{\citenamefont {Yi}\ \emph {et~al.}(2014)\citenamefont {Yi}, \citenamefont {Zhang}, \citenamefont {Liu}, \citenamefont {Ding}, \citenamefont {Chu}, \citenamefont {Kemper}, \citenamefont {Plonka}, \citenamefont {Moritz}, \citenamefont {Hashimoto}, \citenamefont {Mo} \emph {et~al.}}]{yi2014dynamic}%
  \BibitemOpen
  \bibfield  {author} {\bibinfo {author} {\bibfnamefont {M.}~\bibnamefont {Yi}}, \bibinfo {author} {\bibfnamefont {Y.}~\bibnamefont {Zhang}}, \bibinfo {author} {\bibfnamefont {Z.-K.}\ \bibnamefont {Liu}}, \bibinfo {author} {\bibfnamefont {X.}~\bibnamefont {Ding}}, \bibinfo {author} {\bibfnamefont {J.-H.}\ \bibnamefont {Chu}}, \bibinfo {author} {\bibfnamefont {A.}~\bibnamefont {Kemper}}, \bibinfo {author} {\bibfnamefont {N.}~\bibnamefont {Plonka}}, \bibinfo {author} {\bibfnamefont {B.}~\bibnamefont {Moritz}}, \bibinfo {author} {\bibfnamefont {M.}~\bibnamefont {Hashimoto}}, \bibinfo {author} {\bibfnamefont {S.-K.}\ \bibnamefont {Mo}}, \emph {et~al.},\ }\bibfield  {title} {\bibinfo {title} {\textit{Dynamic competition between spin-density wave order and superconductivity in underdoped} {Ba$_{(1-x)}$K$_x$Fe$_2$$As_2$}},\ }\href@noop {} {\bibfield  {journal} {\bibinfo  {journal} {Nat. Commun.}\ }\textbf {\bibinfo {volume} {5}},\ \bibinfo {pages} {3711} (\bibinfo {year} {2014})}\BibitemShut {NoStop}%
\bibitem [{\citenamefont {Fernandes}\ \emph {et~al.}(2014)\citenamefont {Fernandes}, \citenamefont {Chubukov},\ and\ \citenamefont {Schmalian}}]{fernandes2014drives}%
  \BibitemOpen
  \bibfield  {author} {\bibinfo {author} {\bibfnamefont {R.}~\bibnamefont {Fernandes}}, \bibinfo {author} {\bibfnamefont {A.}~\bibnamefont {Chubukov}},\ and\ \bibinfo {author} {\bibfnamefont {J.}~\bibnamefont {Schmalian}},\ }\bibfield  {title} {\bibinfo {title} {\textit{What drives nematic order in iron-based superconductors?}},\ }\href@noop {} {\bibfield  {journal} {\bibinfo  {journal} {Nat. Phys.}\ }\textbf {\bibinfo {volume} {10}},\ \bibinfo {pages} {97} (\bibinfo {year} {2014})}\BibitemShut {NoStop}%
\bibitem [{\citenamefont {He}\ \emph {et~al.}(2021)\citenamefont {He}, \citenamefont {Chen}, \citenamefont {Li}, \citenamefont {Zhao}, \citenamefont {Song}, \citenamefont {Yoshida}, \citenamefont {Eisaki}, \citenamefont {Wu}, \citenamefont {Chen}, \citenamefont {Lu} \emph {et~al.}}]{he2021superconducting}%
  \BibitemOpen
  \bibfield  {author} {\bibinfo {author} {\bibfnamefont {Y.}~\bibnamefont {He}}, \bibinfo {author} {\bibfnamefont {S.-D.}\ \bibnamefont {Chen}}, \bibinfo {author} {\bibfnamefont {Z.-X.}\ \bibnamefont {Li}}, \bibinfo {author} {\bibfnamefont {D.}~\bibnamefont {Zhao}}, \bibinfo {author} {\bibfnamefont {D.}~\bibnamefont {Song}}, \bibinfo {author} {\bibfnamefont {Y.}~\bibnamefont {Yoshida}}, \bibinfo {author} {\bibfnamefont {H.}~\bibnamefont {Eisaki}}, \bibinfo {author} {\bibfnamefont {T.}~\bibnamefont {Wu}}, \bibinfo {author} {\bibfnamefont {X.-H.}\ \bibnamefont {Chen}}, \bibinfo {author} {\bibfnamefont {D.-H.}\ \bibnamefont {Lu}}, \emph {et~al.},\ }\bibfield  {title} {\bibinfo {title} {\textit{Superconducting fluctuations in overdoped} {Bi$_2$Sr$_2$CaCu$_2$O$_{8+\delta}$}},\ }\href@noop {} {\bibfield  {journal} {\bibinfo  {journal} {Phys. Rev. X}\ }\textbf {\bibinfo {volume} {11}},\ \bibinfo {pages} {031068} (\bibinfo {year} {2021})}\BibitemShut {NoStop}%
\bibitem [{\citenamefont {Giannozzi}\ \emph {et~al.}(2009)\citenamefont {Giannozzi}, \citenamefont {Baroni}, \citenamefont {Bonini}, \citenamefont {Calandra}, \citenamefont {Car}, \citenamefont {Cavazzoni}, \citenamefont {Ceresoli}, \citenamefont {Chiarotti}, \citenamefont {Cococcioni}, \citenamefont {Dabo} \emph {et~al.}}]{giannozzi2009quantum}%
  \BibitemOpen
  \bibfield  {author} {\bibinfo {author} {\bibfnamefont {P.}~\bibnamefont {Giannozzi}}, \bibinfo {author} {\bibfnamefont {S.}~\bibnamefont {Baroni}}, \bibinfo {author} {\bibfnamefont {N.}~\bibnamefont {Bonini}}, \bibinfo {author} {\bibfnamefont {M.}~\bibnamefont {Calandra}}, \bibinfo {author} {\bibfnamefont {R.}~\bibnamefont {Car}}, \bibinfo {author} {\bibfnamefont {C.}~\bibnamefont {Cavazzoni}}, \bibinfo {author} {\bibfnamefont {D.}~\bibnamefont {Ceresoli}}, \bibinfo {author} {\bibfnamefont {G.~L.}\ \bibnamefont {Chiarotti}}, \bibinfo {author} {\bibfnamefont {M.}~\bibnamefont {Cococcioni}}, \bibinfo {author} {\bibfnamefont {I.}~\bibnamefont {Dabo}}, \emph {et~al.},\ }\bibfield  {title} {\bibinfo {title} {\textit{{QUANTUM ESPRESSO}: A Modular and open-source software project for quantum simulations of materials}},\ }\href@noop {} {\bibfield  {journal} {\bibinfo  {journal} {J. Condens. Matter Phys.}\ }\textbf {\bibinfo {volume} {21}},\ \bibinfo {pages} {395502} (\bibinfo {year} {2009})}\BibitemShut {NoStop}%
\bibitem [{\citenamefont {Giannozzi}\ \emph {et~al.}(2017)\citenamefont {Giannozzi}, \citenamefont {Andreussi}, \citenamefont {Brumme}, \citenamefont {Bunau}, \citenamefont {Nardelli}, \citenamefont {Calandra}, \citenamefont {Car}, \citenamefont {Cavazzoni}, \citenamefont {Ceresoli}, \citenamefont {Cococcioni} \emph {et~al.}}]{giannozzi2017advanced}%
  \BibitemOpen
  \bibfield  {author} {\bibinfo {author} {\bibfnamefont {P.}~\bibnamefont {Giannozzi}}, \bibinfo {author} {\bibfnamefont {O.}~\bibnamefont {Andreussi}}, \bibinfo {author} {\bibfnamefont {T.}~\bibnamefont {Brumme}}, \bibinfo {author} {\bibfnamefont {O.}~\bibnamefont {Bunau}}, \bibinfo {author} {\bibfnamefont {M.~B.}\ \bibnamefont {Nardelli}}, \bibinfo {author} {\bibfnamefont {M.}~\bibnamefont {Calandra}}, \bibinfo {author} {\bibfnamefont {R.}~\bibnamefont {Car}}, \bibinfo {author} {\bibfnamefont {C.}~\bibnamefont {Cavazzoni}}, \bibinfo {author} {\bibfnamefont {D.}~\bibnamefont {Ceresoli}}, \bibinfo {author} {\bibfnamefont {M.}~\bibnamefont {Cococcioni}}, \emph {et~al.},\ }\bibfield  {title} {\bibinfo {title} {\textit{Advanced capabilities for materials modelling with {Quantum ESPRESSO}}},\ }\href@noop {} {\bibfield  {journal} {\bibinfo  {journal} {J. Condens. Matter Phys.}\ }\textbf {\bibinfo {volume} {29}},\ \bibinfo {pages} {465901} (\bibinfo {year} {2017})}\BibitemShut {NoStop}%
\bibitem [{\citenamefont {Furness}\ \emph {et~al.}(2020)\citenamefont {Furness}, \citenamefont {Kaplan}, \citenamefont {Ning}, \citenamefont {Perdew},\ and\ \citenamefont {Sun}}]{furness2020accurate}%
  \BibitemOpen
  \bibfield  {author} {\bibinfo {author} {\bibfnamefont {J.~W.}\ \bibnamefont {Furness}}, \bibinfo {author} {\bibfnamefont {A.~D.}\ \bibnamefont {Kaplan}}, \bibinfo {author} {\bibfnamefont {J.}~\bibnamefont {Ning}}, \bibinfo {author} {\bibfnamefont {J.~P.}\ \bibnamefont {Perdew}},\ and\ \bibinfo {author} {\bibfnamefont {J.}~\bibnamefont {Sun}},\ }\bibfield  {title} {\bibinfo {title} {\textit{Accurate and numerically efficient {r$^2$SCAN} meta-generalized gradient approximation}},\ }\href@noop {} {\bibfield  {journal} {\bibinfo  {journal} {J. Phys. Chem.}\ }\textbf {\bibinfo {volume} {11}},\ \bibinfo {pages} {8208} (\bibinfo {year} {2020})}\BibitemShut {NoStop}%
\bibitem [{\citenamefont {Grimme}(2006)}]{grimme2006semiempirical}%
  \BibitemOpen
  \bibfield  {author} {\bibinfo {author} {\bibfnamefont {S.}~\bibnamefont {Grimme}},\ }\bibfield  {title} {\bibinfo {title} {\textit{Semiempirical GGA-type density functional constructed with a long-range dispersion correction}},\ }\href@noop {} {\bibfield  {journal} {\bibinfo  {journal} {J. Comput. Chem.}\ }\textbf {\bibinfo {volume} {27}},\ \bibinfo {pages} {1787} (\bibinfo {year} {2006})}\BibitemShut {NoStop}%
\bibitem [{\citenamefont {Kaneko}\ \emph {et~al.}(2013)\citenamefont {Kaneko}, \citenamefont {Toriyama}, \citenamefont {Konishi},\ and\ \citenamefont {Ohta}}]{kaneko2013orthorhombic}%
  \BibitemOpen
  \bibfield  {author} {\bibinfo {author} {\bibfnamefont {T.}~\bibnamefont {Kaneko}}, \bibinfo {author} {\bibfnamefont {T.}~\bibnamefont {Toriyama}}, \bibinfo {author} {\bibfnamefont {T.}~\bibnamefont {Konishi}},\ and\ \bibinfo {author} {\bibfnamefont {Y.}~\bibnamefont {Ohta}},\ }\bibfield  {title} {\bibinfo {title} {\textit{Orthorhombic-to-monoclinic phase transition of} {Ta$_2$NiSe$_5$} \textit{induced by the Bose-Einstein condensation of excitons}},\ }\href@noop {} {\bibfield  {journal} {\bibinfo  {journal} {Phys. Rev. B}\ }\textbf {\bibinfo {volume} {87}},\ \bibinfo {pages} {035121} (\bibinfo {year} {2013})}\BibitemShut {NoStop}%
\bibitem [{fin()}]{finiteSize}%
  \BibitemOpen
  \href@noop {} {}\bibinfo {note} {In a finite-size system, the exact ground state does not spontaneously break symmetries, so that $\langle \hat{x}_{\rm ph}\rangle\equiv 0$.}\BibitemShut {Stop}%
\end{thebibliography}%

\title{Supplementary Information \\Anomalous excitonic phase diagram in band-gap-tuned Ta$_2$Ni(Se,S)$_5$}

\author{Cheng Chen}
\email{These authors contributed equally to this work}
\affiliation{Department of Physics, University of Oxford, Oxford, OX1 3PU, United Kingdom}
\affiliation{Department of Applied Physics, Yale University, New Haven, Connecticut 06511, USA}

\author{Weichen Tang} 
\email{These authors contributed equally to this work}
\affiliation{Physics Department, University of California, Berkeley, California 94720, USA}
\affiliation{Materials Science Division, Lawrence Berkeley National Lab, Berkeley, California 94720, USA}

\author{Xiang Chen} 
\affiliation{Physics Department, University of California, Berkeley, California 94720, USA}
\affiliation{Materials Science Division, Lawrence Berkeley National Lab, Berkeley, California 94720, USA}

\author{Zhibo Kang}
\affiliation{Department of Applied Physics, Yale University, New Haven, Connecticut 06511, USA}

\author{Shuhan Ding} 
\affiliation{Department of Physics and Astronomy, Clemson University, Clemson, South Carolina 29631, USA}

\author{Kirsty Scott} 
\affiliation{Department of Applied Physics, Yale University, New Haven, Connecticut 06511, USA}

\author{Siqi Wang} 
\affiliation{Department of Applied Physics, Yale University, New Haven, Connecticut 06511, USA}

\author{Zhenglu Li} 
\affiliation{Physics Department, University of California, Berkeley, California 94720, USA}
\affiliation{Materials Science Division, Lawrence Berkeley National Lab, Berkeley, California 94720, USA}

\author{Jacob P.C. Ruff}
\affiliation{Cornell High Energy Synchrotron Source, Cornell University, Ithaca, New York 14853, USA}

\author{Makoto Hashimoto}
\affiliation{Stanford Synchrotron Radiation Lightsource, SLAC National Accelerator Laboratory, Menlo Park, California 94025, USA}

\author{Dong-Hui~Lu}
\affiliation{Stanford Synchrotron Radiation Lightsource, SLAC National Accelerator Laboratory, Menlo Park, California 94025, USA}

\author{Chris Jozwiak}
\affiliation{Advanced Light Source, Lawrence Berkeley National Laboratory, Berkeley, California 94720, USA}

\author{Aaron Bostwick}
\affiliation{Advanced Light Source, Lawrence Berkeley National Laboratory, Berkeley, California 94720, USA}

\author{Eli Rotenberg}
\affiliation{Advanced Light Source, Lawrence Berkeley National Laboratory, Berkeley, California 94720, USA}

\author{Eduardo H. da Silva Neto}
\affiliation{Department of Physics, Yale University, New Haven, Connecticut 06511, USA}

\author{Robert J. Birgeneau}
\affiliation{Physics Department, University of California, Berkeley, California 94720, USA}
\affiliation{Materials Science Division, Lawrence Berkeley National Lab, Berkeley, California 94720, USA}
\affiliation{Department of Materials Science and Engineering, University of California, Berkeley, California 94720, USA}

\author{Yulin Chen}
\affiliation{Department of Physics, University of Oxford, Oxford, OX1 3PU, United Kingdom}

\author{Steven G. Louie} 
\email{sglouie@berkeley.edu}
\affiliation{Physics Department, University of California, Berkeley, California 94720, USA}
\affiliation{Materials Science Division, Lawrence Berkeley National Lab, Berkeley, California 94720, USA}
\affiliation{Department of Materials Science and Engineering, University of California, Berkeley, California 94720, USA}

\author{Yao Wang} 
\email{yao.wang@emory.edu}
\affiliation{Department of Physics and Astronomy, Clemson University, Clemson, South Carolina 29631, USA}
\affiliation{Department of Chemistry, Emory University, Atlanta, GA 30322, USA}

\author{Yu He}
\email{yu.he@yale.edu}
\affiliation{Department of Applied Physics, Yale University, New Haven, Connecticut 06511, USA}
\affiliation{Physics Department, University of California, Berkeley, California 94720, USA}
\affiliation{Materials Science Division, Lawrence Berkeley National Lab, Berkeley, California 94720, USA}

\date{\today}

\maketitle

\makeatletter
\def\fnum@figure{\figurename\thefigure}
\makeatother

\setcounter{figure}{0}
\renewcommand{\figurename}{Fig.~S}

\clearpage

\section{Supplementary Note 1: Complete data of resistivity, XRD, ARPES and EDX measurements}

\begin{figure*}[!hb]
\centering
\includegraphics[width= 16cm]{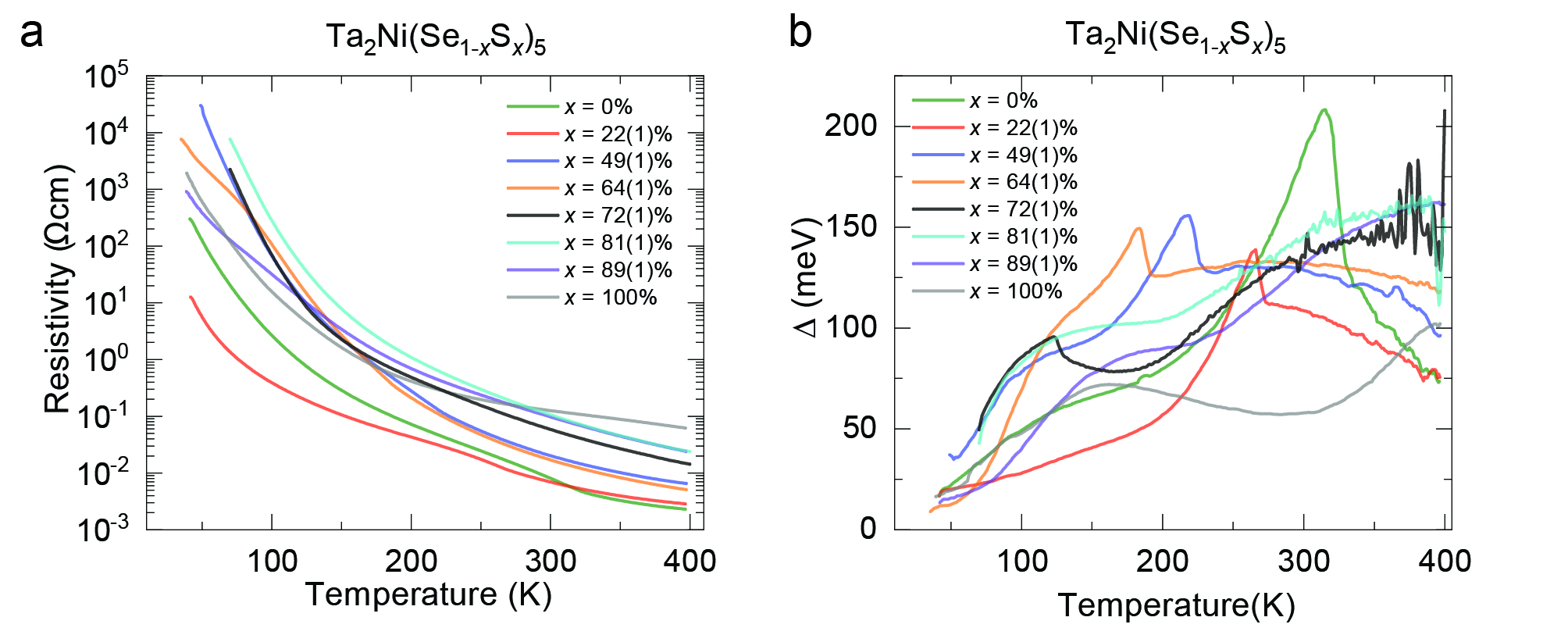}
\caption{\textbf{Resistivity measurement of \TNSS.} \textbf{a} Temperature-dependent resistivity of \TNSS. \textbf{b} The temperature dependence of the activation energy $\Delta$ of the resistivity $\rho$ in \textbf{a} given by $\Delta=-k_B T^2 (\partial{\ln\rho}/\partial{T})$. The actual S doping level $x$ of each sample is determined by energy-dispersive x-ray spectroscopy, in which the typical variation of $x$ is less than 1$\%$.}
\label{fig:SFig1}
\end{figure*}

Resistivity measurements were performed on each \TNSS~ sample with different S-doping levels, and the result is illustrated in Fig.~S\ref{fig:SFig1}\textbf{a}. Similar temperature evolution is evidenced as the sample becomes more insulating towards low temperatures. However, the resistivity curve does not show clear signs of metal-to-insulator transition. The structural symmetry-breaking phase transition only manifests into a small kink, for instance around $T_s=$ 329~K in the pristine \TNSe. This feature can be better visualized as the peak in activation energy $\Delta$ of the resistivity $\rho$, shown in Fig.~S\ref{fig:SFig1}\textbf{b}. Such peaks are only evidenced in samples with S-doping levels below $72\%$ (result plotted in Fig. 2\textbf{b} of the main text). Therefore, we note that transport measurement is not a direct measurement of the structural phase transition.

\begin{figure*}
\centering
\includegraphics[width= 14cm]{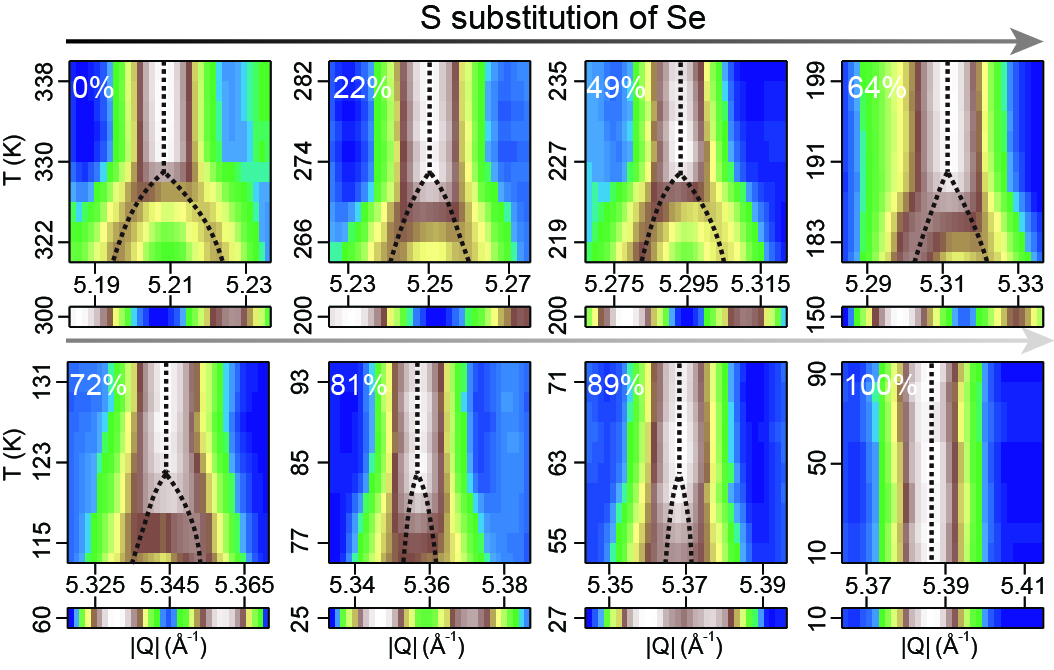}
\caption{\textbf{High energy X-ray diffraction data of \TNSS.} Evolution of the Bragg peak $\pm$ 2 4 8 as a function of temperature for samples with different S-doping levels. The splitting of the peak marks the structural phase transition from the high-temperature orthorhombic phase to the low-temperature monoclinic phase. No structural transition is observed in \TNS~ within the resolution limit of the experiment (about 0.01 \AA$^{-1}$). Black dashed lines are guides for the eyes of the Bragg peak positions.  }
\label{fig:SFig2}
\end{figure*}

Complete data for high-resolution synchrotron-based XRD measurement on \TNSS~ family is illustrated in Fig.~S\ref{fig:SFig2}. Crossing the second-order structural phase transition, the system turns from the high-temperature orthorhombic phase to the low-temperature monoclinic phase, and some Bragg peaks, for instance the 2 4 8 Bragg peak shown here, split into two. Such splitting is evidenced in all samples with different S-doping levels, except for \TNS~, where no structural transition is observed within the resolution limit of the experiment. The temperature where the splitting starts marks the structural phase transition point deduced from XRD measurement and the result is summarized in Fig. 2\textbf{a} of the main text. On the other side, the evolution of $\beta$ angle, i.e. structural order parameter in the low-temperature monoclinic phase, can be deduced from the separation of the split peak, and the result is plotted in Fig. 2\textbf{b} of the main text.

\begin{figure*}
\centering
\includegraphics[width= 7cm]{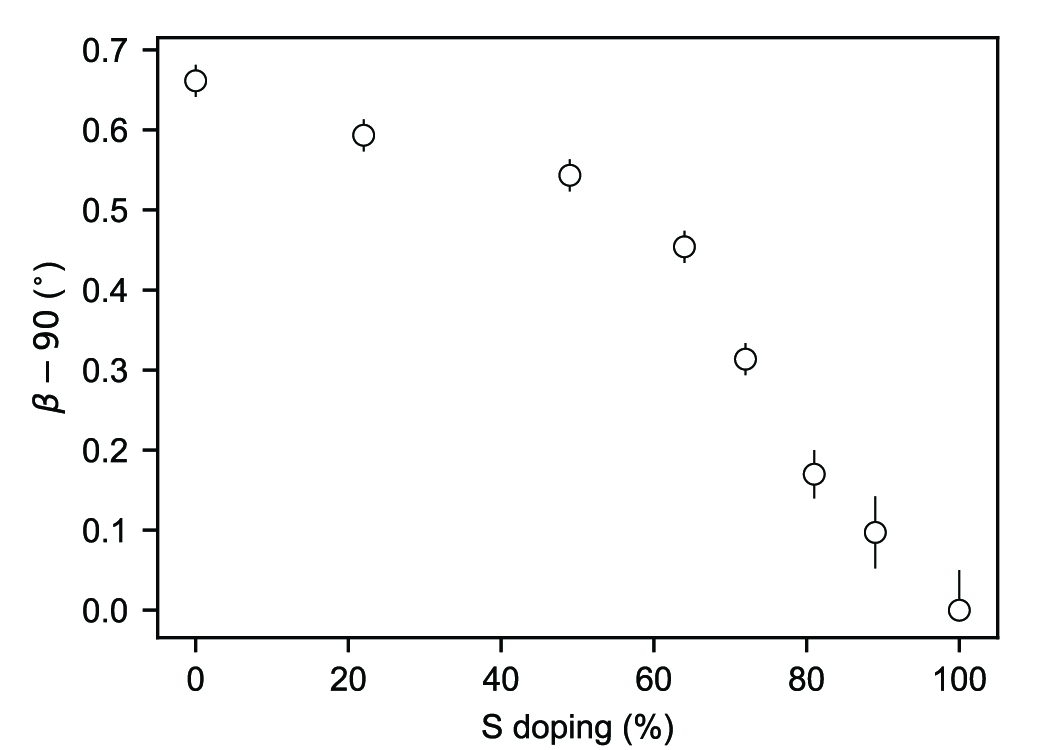}
\caption{\textbf{Structural order parameter at the temperature of ARPES measurement} The amplitude of the deviation of the $\beta$ angle from $90^\circ$ at the temperature of ARPES measurement, as compared to the maximum value. No observable structural phase transition is observed at 100$\%$ S doped compound (Ta$_2$NiS$_5$). }
\label{fig:SFig3}
\end{figure*}

\begin{figure*}
\centering
\includegraphics[width= 17cm]{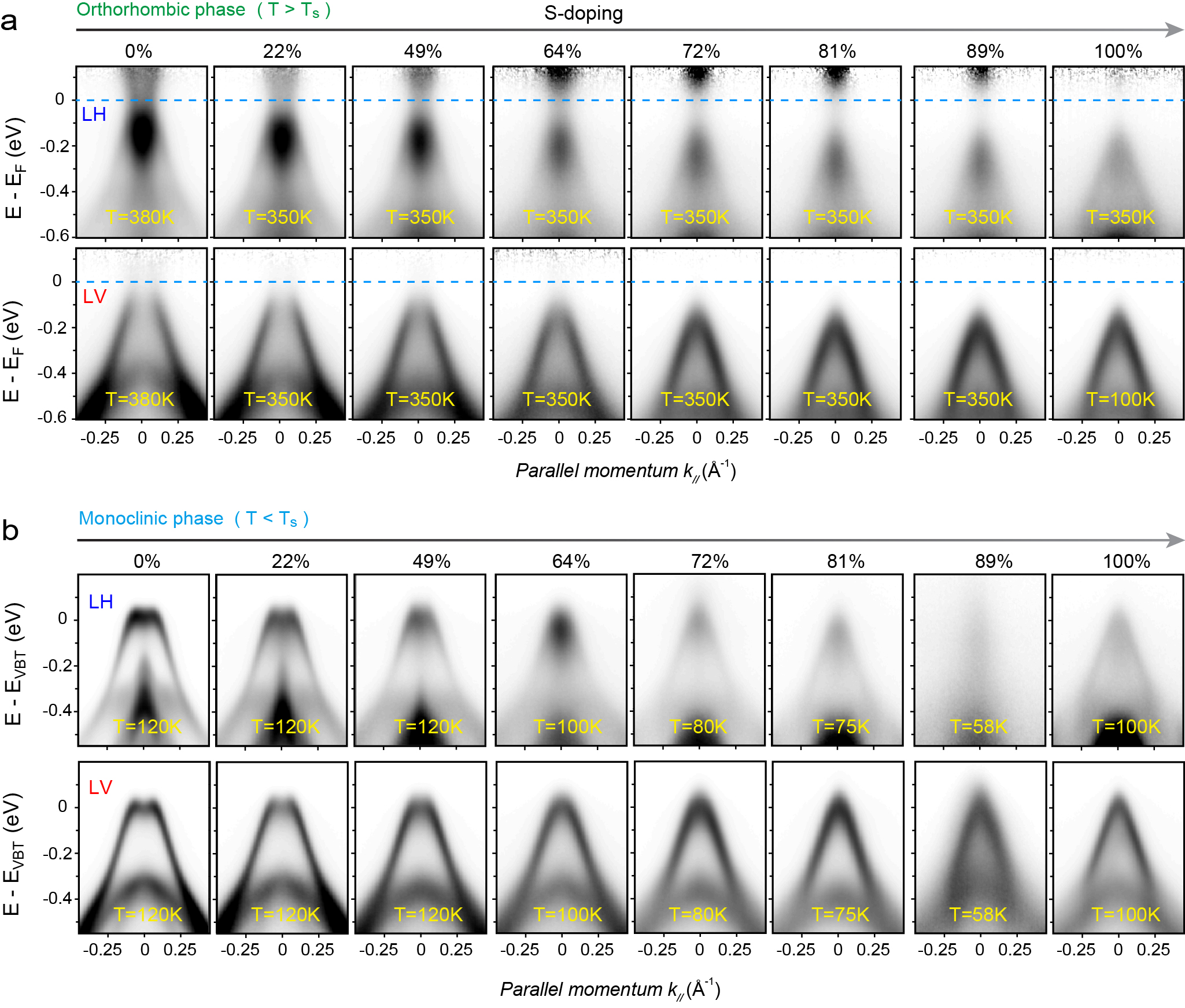}
\caption{\textbf{Photoemission data of \TNSS.} \textbf{a} Photoemission spectra along $X-\Gamma-X$ direction of \TNSS~ taken in high-temperature orthorhombic phase. LH (linear horizontal) and LV (linear vertical) denote the polarization of the incident photon beam. \textbf{b} Same as \textbf{a} but for spectra taken in low-temperature monoclinic phase, except for the 100$\%$ sulfur substituted compound (Ta$_2$NiS$_5$) where no sign of a structural phase transition is observed in the XRD measurement. }
\label{fig:SFig4}
\end{figure*}

High statistics and energy resolution ARPES measurements are performed on \TNSS~ family at both the high-temperature orthorhombic phase and the low-temperature monoclinic phase (except for at 100$\%$ S-doping where no structural phase transition is observed) to trace the evolution of electronic structure with S-doping. Since the structural phase transition is second order in nature, we plot the structural order parameter at the temperature of ARPES measurement in Fig.~S\ref{fig:SFig3}, to ensure that visible features of the monoclinic phase have already developed. Linear horizontal (LH) and linear vertical (LV) incident beams from synchrotron radiation are employed, selectively probing bands with different orbital origins, which in this case highlight the conduction and valence bands respectively. The complete data is illustrated in Fig.~S\ref{fig:SFig4}. For the high-temperature data, we divide the resolution-convolved Fermi-Dirac function to restore the spectra up to $\sim$150 meV above the Fermi level. For the low-temperature data, as the samples become insulating and the Fermi level is not pinned at a well-defined position, we aligned the spectra of different samples to the top of the valence band to ensure a direct comparison. For better visualization of the ARPES result, we combined the intensity of LH and LV channels with equal contributions in blue and red colors and plotted in Fig. 3 of the main text.

\begin{figure*}
\centering
\includegraphics[width= 14cm]{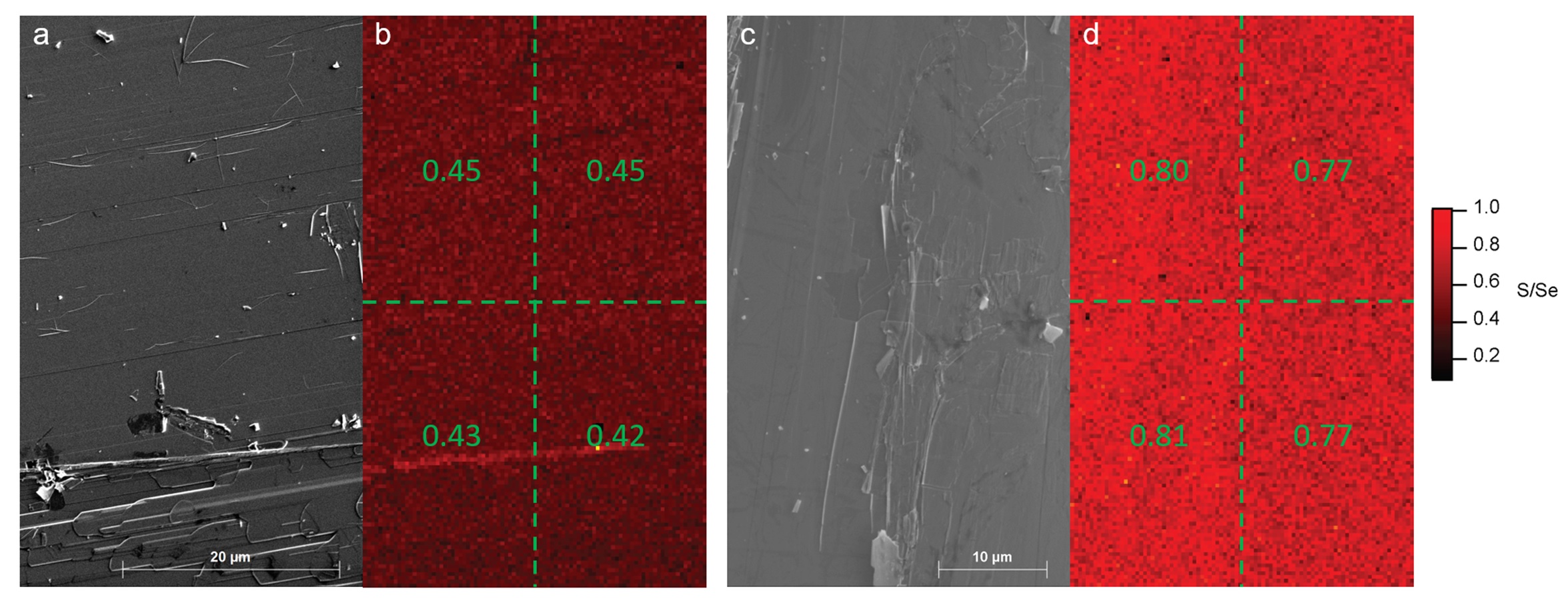}
\caption{\textbf{Spatial homogeneity of S doping in \TNSS.} \textbf{a(c)} SEM image of the surface of a sample with 64$\%$ and 72$\%$ S-doping. \textbf{b(d)} The false-color plot of the ratio between the intensity of sulfur K-alpha peak and selenium L peak corresponding to a(c). The green numbers on b(d) indicate the ratio between the intensity of sulfur K-alpha peak and selenium L peak averaged in the quarters segmented by the green dashed lines. The variation of the actual S-doping level determined from our EDX measurements in each sample is typically less than 1$\%$.}
\label{fig:SI_EDX}
\end{figure*}

To check the spatial homogeneity of sulfur doping within each doped sample, we performed energy dispersive X-ray spectroscopy (EDX) mapping over the sample surface, and Fig.~S\ref{fig:SI_EDX} shows the results on the select samples with 64$\%$ and 72$\%$ S-doping levels. We find that the S-doping level within each sample is homogeneous with an average variation of up to $\pm2\%$ between regions of 10-20 microns lateral dimensions, which matches our smallest ARPES beam spot sizes.

\clearpage

\section{Supplementary Note 2: Fitting of photoemission spectra}

\begin{figure*}
\centering
\includegraphics[width= 10cm]{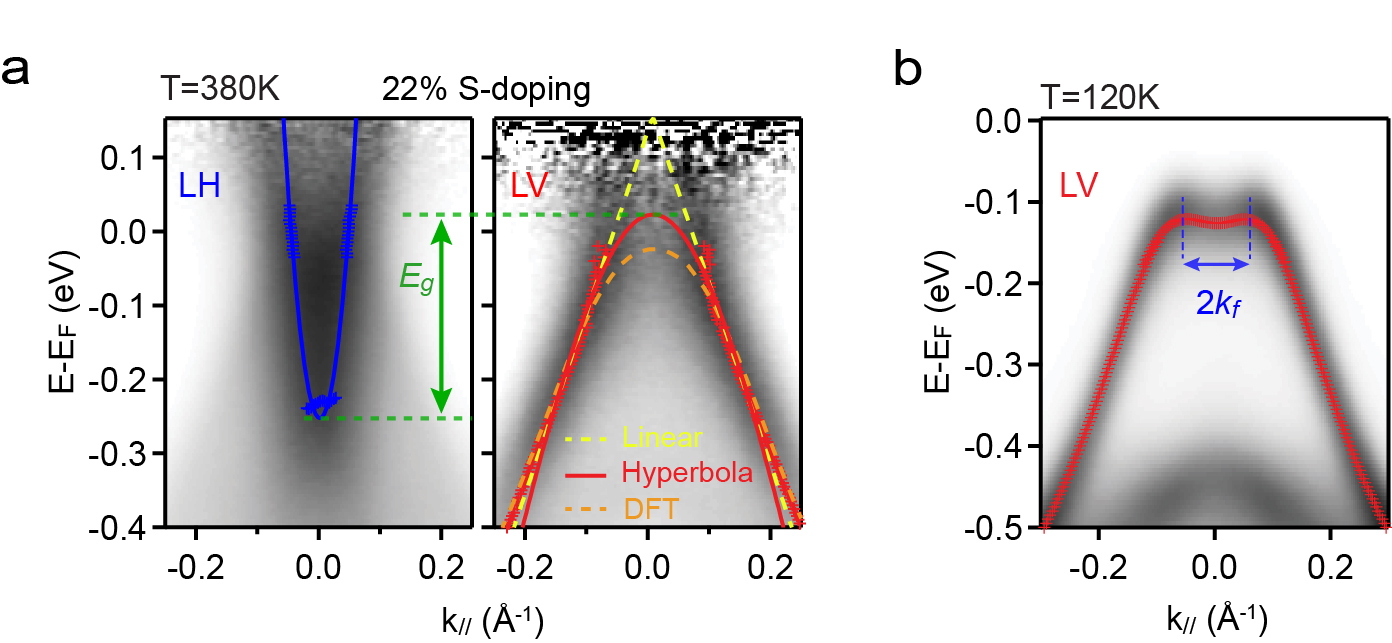}
\caption{\textbf{Photoemission spectra fitting.} \textbf{a} High-temperature photoemission spectra of a 22$\%$ S-doped sample in both LH (linear horizontal) and LV (linear vertical) channels, highlighting conduction and valence band respectively. The spectra are normalized along the energy axis to remove the effect of the pseudogap state \cite{chen2022lattice}. The “+” marks are deduced from the peak fitting of energy distribution curves (EDCs) or momentum distribution curves (MDCs), which were then fitted by hyperbola to extract the size of the band overlap $E_g$. The upper (lower) bound of $E_g$ is estimated from the fitting of the valence band with a linear (normalized DFT dispersion) band shape. \textbf{b} Low-temperature photoemission spectra of the 22$\%$ S-doped sample in LV channel. Band dispersion is deduced from the peak fitting of EDCs or MDCs, from which the size of $2k_F$ is extracted. }
\label{fig:SFig6}
\end{figure*}

To estimate the band overlap/gap $E_g$ in the high-temperature ARPES spectra, we fitted both the conduction and valence band dispersion. As illustrated in Fig.~S\ref{fig:SFig6}, the spectra were first normalized along the energy direction to compensate for the spectra weight depletion around the Fermi level, resulting from the pseudogap state\cite{chen2022lattice}. Then, the dispersion of the bands was traced by peak-fitting the energy distribution curves (EDCs) or momentum distribution curves (MDCs) of the ARPES spectra. The conduction band is fitted with the parabolic curve (blue dashed line), and the valence band is fitted using 3 different band shapes: linear (dashed yellow line), hyperbola (red line), and DFT result (dashed orange line). The upper and lower bound of $E_g$ comes from the fitting result of the linear and DFT band shapes, respectively. In the case of quantifying low-temperature band back-bending momentum $k_F$, similarly, we trace the dispersion of the valence band by peak-fitting the EDCs and MDCs of the ARPES spectra. The 2$k_F$ value is taken from the distance between the valence band top and the error comes from the momentum resolution of the ARPES measurement.

\clearpage

\section{Supplementary Note 3: DFT calculation}

\begin{figure*}
\centering
\includegraphics[width= 17cm]{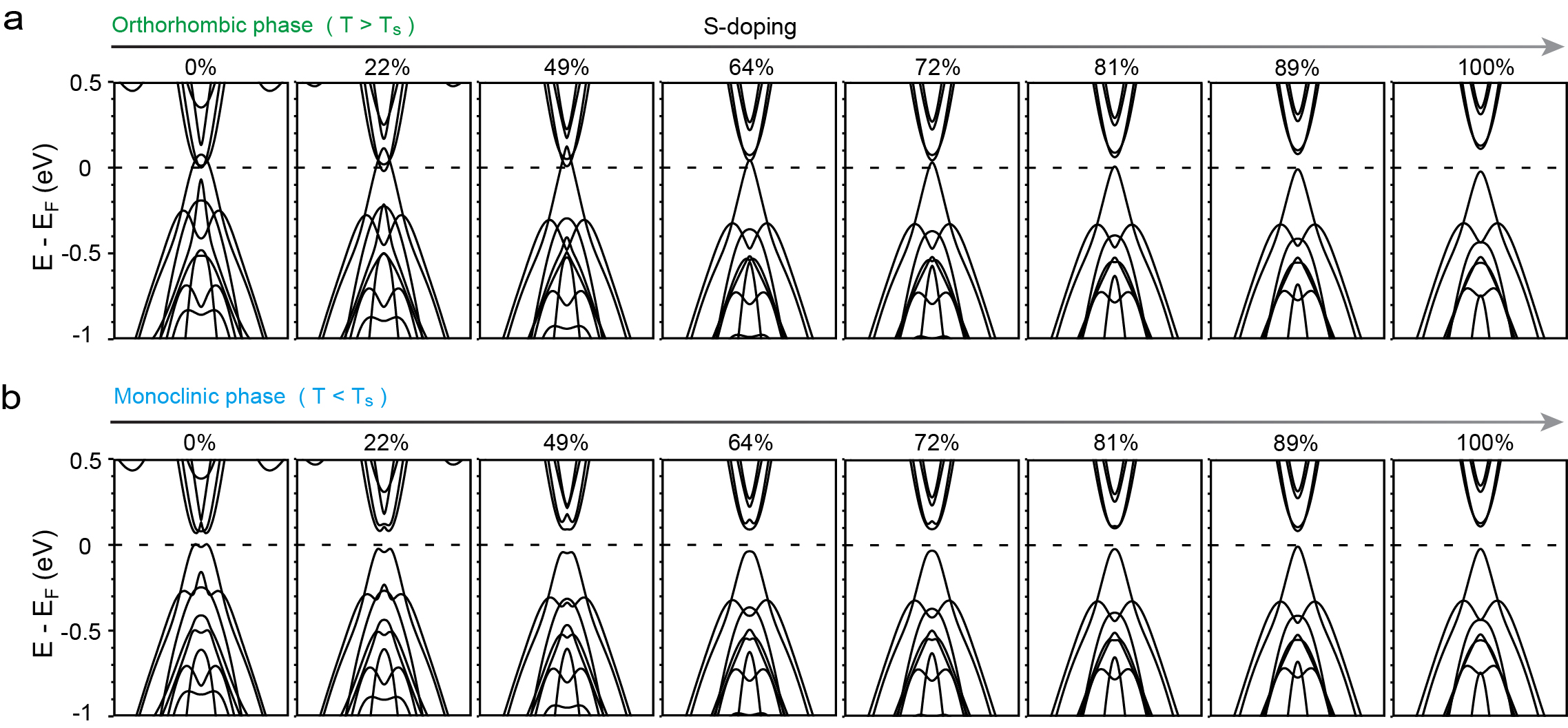}
\caption{\textbf{DFT calculation of \TNSS.} Calculated band structure along $X-\Gamma-X$ direction of \TNSS~ in \textbf{a} orthorhombic lattice structure, \textbf{b} monoclinic lattice structure, respectively.}
\label{fig:SFig5}
\end{figure*}

Figure~S\ref{fig:SFig5} shows complete data of DFT calculation on \TNSS~ in both the orthorhombic and the monoclinic phases. The orthorhombic phase is achieved by imposing the system's symmetry to conform to the orthorhombic structure, while the monoclinic phase is attained through full relaxation of the structure. The result is overlayed on the ARPES spectra in Fig. 3\textbf{a} of the main text. The band overlap/gap $E_g$ in the orthorhombic phase and the band back back-bending momentum $k_F$ in the monoclinic phase are also deduced and the result is plotted in Fig.~3\textbf{c-d} of the main text.

\clearpage

\section{Supplementary Note 4: Impact of the Coulomb Interaction}

\begin{figure*}[!t]
\centering
\includegraphics[width= 14cm]{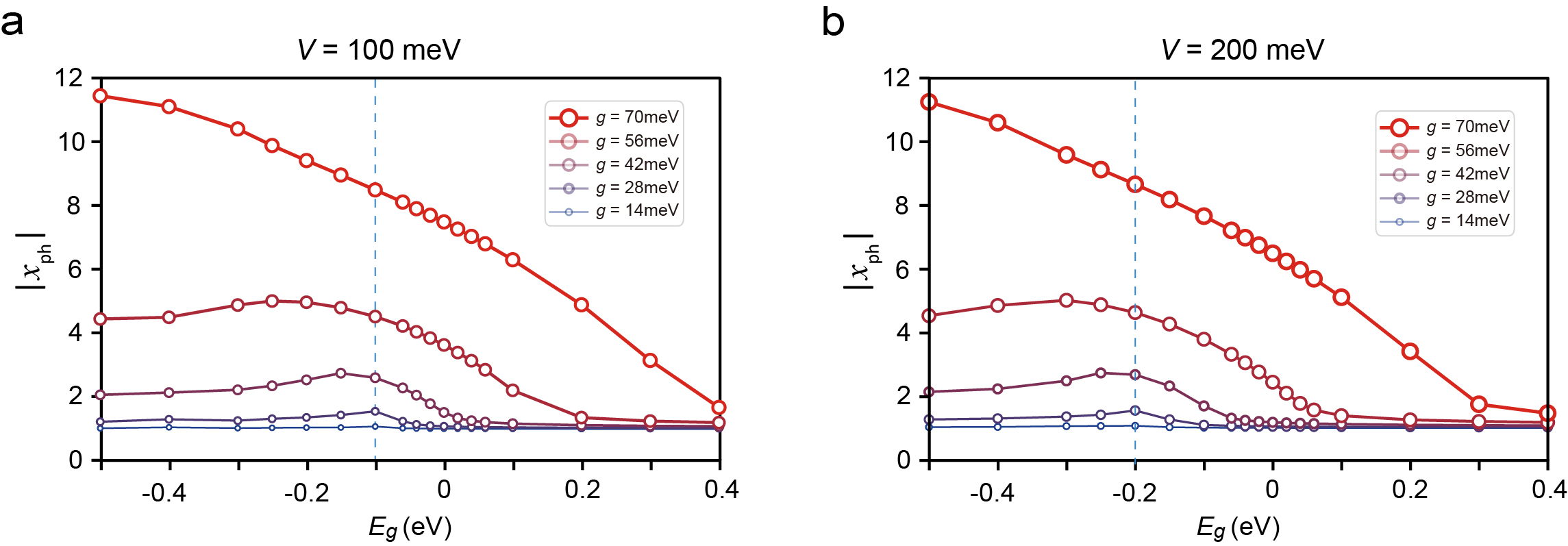}\vspace{-5mm}
\caption{\textbf{Influence of direct Coulomb interaction $V$.} The calculated mean-average lattice displacement $|x_{\rm ph}|$ as a function of the band gap $E_g$ at various electron-phonon coupling strengths, with finite Coulomb interaction (a) $V=100$\,meV. and (b) $V=200$\,meV. (The $V=0$ case is shown in Fig. 4 of main text).  }
\label{fig:SFig7}
\end{figure*}

To understand the impact of Coulomb interactions [the $V$ in Eq.~(1) of the main text], we incorporate non-zero interaction terms into the simulation of the many-body system. Figure~S\ref{fig:SFig7} shows the mean-average lattice displacement $|x_{\rm ph}|$ evaluated using ED simulations for systems with $V=100$\,meV and 200\,meV. While considering the upper limit of the experimentally determined Coulomb interaction around 70\,meV\cite{chen2022lattice}, we allow for adjustments within a certain range to accommodate potential variations induced by S-doping. Compared to the Fig.~4 of the main text, the dependence of lattice displacement on $E_g$ presented in Fig. S\ref{fig:SFig7} shows similar behavior with an overall shift equal to the Coulomb interaction $V$. Such a shift can be better characterized by the hump positions in the weak-coupling limit. The overall move of the critical $E_g$ reflects a relative motion between two bands caused by the Coulomb interaction $V$ and this phenomenon can be attributed to the Hartree component.

\begin{figure*}[!t]
\centering
\includegraphics[width= 14cm]{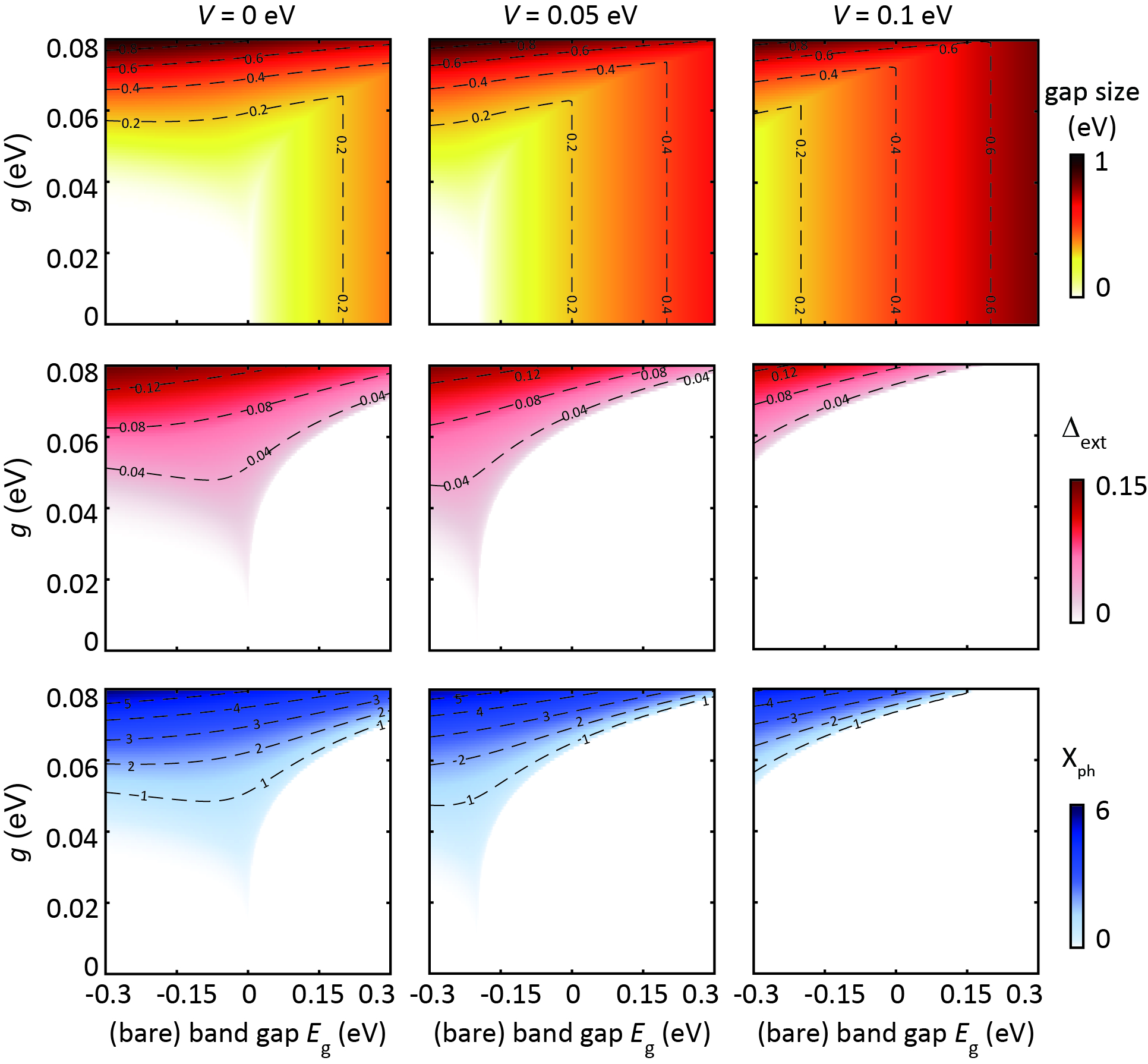}
\caption{\textbf{Mean-field solutions for various interactions and bare band gap sizes.} Upper panels: Direct mean-field gap as a function of electron-phonon coupling $g$ and bare band gap $E_g$, for interacting electrons with $V=0$ (left), $V=0.05$\,eV (middle), and $V=0.1$\,eV (right). Middle and lower panels: Mean-field order parameter for electrons $\Delta_{\rm ext}$ (red) and lattice $X_{\rm ph}$ (blue). }
\label{fig:SFig8}
\end{figure*}

In order to gain deeper insights into the influence of the parameter $V$, we perform a mean-field (MF) analysis on the many-body Hamiltonian [Eq.~(1) of the main text]. Specifically, we consider the MF excitonic order parameter $\Delta_{\rm ext}=\sum_{i\sigma}\langle f^\dagger_{i\sigma} c_{i\sigma}\rangle/2N$ and $\Delta^\prime_{\rm ext} = \sum_{i\sigma}\langle f^\dagger_{i\sigma}c_{i+1,\sigma}\rangle/2N$, which are identical when the reflection symmetry is preserved, and the MF lattice order parameter $X_{\rm ph}=\langle x_{\rm ph}\rangle$. The MF decomposition of Eq.~(1) is as follows
\begin{eqnarray}
\mathcal{H}_{\rm MF} &=& \sum_{k\sigma} (c_{k\sigma}^\dagger,\, f_{k\sigma}^\dagger)
    \left(\begin{array}{cc}
      \varepsilon^c_k + 2V\sum_{i\sigma'}\langle n^f_{i\sigma'} \rangle/N   & gX_{\rm ph} - V(\Delta_{\rm ext} + {\rm e}^{-{\rm i}k}\Delta^\prime_{\rm ext}) \\
       gX_{\rm ph} - V(\Delta_{\rm ext} + {\rm e}^{-{\rm i}k}\Delta^\prime_{\rm ext})   &  \varepsilon^v_k + 2V\sum_{i\sigma'}\langle n^c_{i\sigma'} \rangle/N
    \end{array}\right)
    \left(\begin{array}{cc}
      c_{k\sigma} \\ f_{k\sigma} \end{array}\right)\nonumber\\
    &&-\sum_{i\sigma\sigma'}\langle n^c_{i\sigma} \rangle\langle n^f_{i\sigma} \rangle + V(\Delta_{\rm ext}^2 + \Delta^{\prime\;2}_{\rm ext}) + \frac{1}{2}N\omega_0 X_{\rm ph}^2,
\end{eqnarray}
where the $V$ contributions to the diagonal terms reflect the Hartree shifts between two bands, and $X_{\rm ph}$ is calculated by minimizing the ground-state energy.

Solving the mean-field Hamiltonian yields the single-particle gap, which is jointly determined by the Hartree shift and the hybridization arising from the (off-diagonal) order parameters. As shown in the upper panels of Fig. S\ref{fig:SFig8}, the single-particle gap depends on both the e-ph coupling $g$ and the bard band gap $E_g$. While a positive $E_g$, reflecting the S-doping in \TNSS, enhances the gap, this resultant gap opening is distinct from the hybridization gap induced by symmetry breaking. This distinction can be reflected in the lower panels, where the order parameters $\Delta_{\rm ext}$ and $X_{\rm ph}$ are strongly suppressed by positive $E_g$. Due to this suppression, these order parameters display a dome-like structure centered around $E_g=0$ in the case of a weak-coupling system. When the e-ph coupling is strong, however, the hybridization gap is beyond the influence of $E_g$, resulting in a monotonic reduction of order parameters as $E_g$ increases. This behavior is consistent with the ED simulation depicted in the Fig.~4 of the main text.

When incorporating the Coulomb interaction $V$ into this MF analysis, its contribution to the order parameters is found to be minor, reflected by the slight lowering of iso-intensity curves when contrasting the middle and left panels. Instead, its primary contribution lies in uniformly shifting the complete phase diagram towards smaller $E_g$ values. Therefore, the \TNSS\ transition is primarily dictated by the strong e-ph coupling, as opposed to the electronic Coulomb interactions.

The inefficiency of the Coulomb interaction in forming excitons results from the small Fermi momentum and the opposite parity of the low-energy bands, which has been discovered by earlier experimental and theoretical studies\,\cite{mazza2020nature, watson2020band}. The mismatch of parity completely excludes the purely Coulomb-driven excitonic orders at $k=0$. This explains the diminishment of order parameters for $E_g$, reflecting the reality close to the \TNS\ side of the material. While such an exclusion is no longer exact for finite Fermi momenta, the small $k_F$ in \TNSe\ leads to limited excitonic instability, compared to the substantial Hartree shift caused by the Coulomb interaction\,\cite{chen2022lattice}. In contrast, the intraband $B_{2g}$ phonon connects the two bands directly and breaks the mirror symmetry. Thus, the coupling to lattice distortion exhibits more efficiency in forming excitons.

\end{document}